\documentclass[12pt]{article}

\usepackage{scicite}
\usepackage{epsf}
\usepackage{ulem}
\usepackage{color}

\usepackage[squaren,Gray]{SIunits}
\usepackage{times}

\usepackage{amsmath}
\usepackage{amsfonts}
\usepackage{amssymb}
\usepackage{graphicx}
\usepackage[latin1]{inputenc}
\usepackage{color}
\newcommand{\cora}[1]{\textcolor{black}{#1}}
\newcommand{\coraa}[1]{\textcolor{black}{#1}}
\newcommand{\alex}[1]{\textcolor{black}{#1}}

\def\uout{\mathbf{u}_\mathrm{out}}
\def\upout{\mathbf{u'}_\mathrm{out}}

\def\rout{\mathbf{r}_\mathrm{out}}
\def\rin{\mathbf{r}_\mathrm{in}}
\def\rpin{\mathbf{r'}_\mathrm{in}}

\title{Distortion matrix concept for deep \coraa{optical} imaging \coraa{in scattering media}} 

\author{Amaury Badon${}^{1}$, Victor Barolle${}^{1}$, Kristina Irsch${}^{2,3}$, Albert C. Boccara${}^{1}$,\\ Mathias Fink${}^{1}$, and Alexandre Aubry${}^{1,\ast}$\\
\\
\normalsize{$^1$ Institut Langevin, ESPCI Paris, PSL University, CNRS, 1 rue Jussieu, 75005 Paris, France}
\normalsize{}\\
\normalsize{$^2$Vision Institute / Quinze-Vingts National Eye Hospital}\\
\normalsize{Sorbonne University, CNRS UMR 7210, INSERM U 068, 17 rue Moreau, 75012 Paris, France}\\
\normalsize{$^3$The Wilmer Eye Institute, The Johns Hopkins University School of Medicine, Baltimore, MD, USA}\\
\\
\normalsize{$^\ast$To whom correspondence should be addressed; E-mail:  alexandre.aubry@espci.fr.}\\
\\
\normalsize{\cora{\textbf{Summary Sentence}: The distortion matrix overcomes aberrations and multiple scattering,}}\\ 
\normalsize{\cora{thereby enabling ultra-deep and wide-field optical imaging.}}}

\date{}
\begin{document}
\baselineskip24pt

\maketitle

\begin{abstract}
{In optical imaging, light propagation is affected by the inhomogeneities of the medium. Sample-induced aberrations and multiple scattering can strongly degrade the image resolution and contrast. Based on a dynamic correction of the incident and/or reflected wave-fronts, adaptive optics has been employed to compensate for those aberrations. However, it only applies to spatially-invariant aberrations or to thin aberrating layers. Here, we propose a global and non-invasive approach based on the distortion matrix concept. This matrix basically connects any focusing point of the image with the distorted part of its wave-front in reflection. A \cora{singular value decomposition} of the distortion matrix allows to correct for high-order aberrations and forward multiple scattering over multiple isoplanatic \cora{modes}. Proof-of-concept experiments are performed through biological tissues \cora{including a turbid cornea}. We demonstrate a Strehl ratio enhancement up to 2500 and recover a diffraction-limited resolution until a depth of ten scattering mean free paths.}%, thus breaking the current field-of-view and depth limitations of AO. 
%Several proof-of-concept imaging experiments are performed through biological tissues and an opaque monkey cornea.
%Several proof-of-concept imaging experiments show an enhancement of the Strehl ratio up to 2500 times and an imaging depth of ten scattering mean free paths through an opaque cornea.}
\end{abstract}

%\maketitle
\section*{INTRODUCTION}

For decades, optical microscopy has been a vital tool in biomedical research to observe live specimens with a sub-micron resolution and with minimal invasiveness. Yet, imaging conditions required for such exquisite performances are rarely gathered. For instance, both the resolution and the contrast drop as the imaging depth increases inside a biological tissue. This observation is a consequence of the spatial variations of the specimen's refractive index that distort the wave-front of both the incoming and outgoing light. When these variations exhibit low spatial frequencies we use the term aberrations while scattering describes the effect of the higher spatial variations. Both these effects limit the use of conventional microscopy to shallow depths or to semi-transparent specimens. Imaging deeper requires to simultaneously compensate for these detrimental phenomena. %Despite their common nature, these two effects are usually treated separately and their incomplete correction ends with a limited effect.

To mitigate the aberrations induced by the specimen, the concept of adaptive optics (AO) has been adapted to microscopy from astronomy where it was developed decades ago \cite{babcock,foy1985feasibility}. Indeed, astronomers faced the same impediment as fluctuations in the atmosphere severely distort the wave-front of the light coming from stars and prevent to obtain a diffraction-limited stellar image. Astronomers then proposed to measure these distortions using a wave-front sensor and to counterbalance it with a dynamic programmable element such as deformable mirrors. Following this concept and the development of deformable mirrors with increasing number of elements, AO already demonstrated its benefits in various imaging techniques such as digital holography \cite{thurman2008correction,tippie2010multiple}, confocal microscopy \cite{booth2002adaptive,tao2011adaptive}, two-photon microscopy \cite{debarre2009image,ji2010adaptive,papadopoulos2017scattering,rueckel2006adaptive}, or optical coherence tomography (OCT) \cite{hermann2004adaptive,adie2012computational}. Unfortunately AO methods usually require a guide star \coraa{or are based on an image sharpness metric. Additionally, they}  are limited to a small region called the isoplanatic patch (IP), the area over which the aberrations can be considered as spatially-invariant. Therefore, there is a need to extend the field-of-view of AO methods by tackling the case of multiple IPs. This issue is particularly decisive for deep imaging where IP size becomes extremely tiny: $<$10 $\mu$m beyond a depth of 1 mm \cite{Judkewitz2015}. \cora{Note that multi-conjugate AO can deal with multiple IPs but this is at the price of a much more complex optical set up~\cite{Rigaut2000,Kam2007,Simmonds2013}}.
%However, the application of AO in microscopy is not straightforward for several reasons. First, the measurement of the aberrated wavefront requires the use of a guide star which could be invasive and not practical \cite{horstmeyer2015guidestar}. To circumvent this limitation, iterative AO approaches propose to blindly correct the wavefront by optimizing in a feedback loop a metric of the final image, the contrast for instance. Yet this approach is limited to cases where you have an correct image to start with \cite{muller1974real}. Secondly, in most AO approaches, the wavefront is measured and/or corrected from the pupil plane. This configuration is relevant in the case of spatially invariant aberrations such as the presence of a thin aberrative layer (the atmosphere in astronomical observations for instance) or a refractive index mismatch in microscopy. In the more general case of a spatially variant aberration, a phase mask in the pupil plane will compensate only for a small portion of the field of view (FOV) called an isoplanetic patch (IP). Obtaining an image with a diffraction-limited resolution over the whole FOV implies a wavefront correction for each IP. As there is no practical ways to determine the number and the positions of this patches, AO is usually limited to small FOV. Note that this issue can be partly solved by placing the DM in the plane of the main aberration but this approach is limited to thin layers and can also present practical disadvantages \cite{wu2015numerical,Mertz15}. 

On the other hand, multiple scattering was long thought to be too complex to be compensated. For deep imaging, a gating mechanism is generally used to reject the multiply-scattered photons and capture only the ballistic light. This gating can be spatial~\cite{minsky1955confocal} as in confocal microscopy or temporal~\cite{Hee1993} as in optical coherence tomography, but they are still depth limited as they rely on the exponentially attenuated ballistic light. In a pioneering experiment, Vellekoop and Mosk demonstrated in 2007 the possibility to restore a diffraction-limited spot through a scattering medium by properly shaping the incoming light \cite{Vellekoop}. Subsequently, a matrix approach of light propagation through complex media was developed \cite{popoff2010measuring}. Relying on the measurement of the Green s functions between each pixel of a spatial light modulator (SLM) and of a charge-coupled device (CCD) camera across a scattering medium, the experimental access to the transmission matrix allows taking advantage of multiple scattering for optimal light focusing~\cite{popoff2010measuring} and communication across a diffusive layer~\cite{popoff2010image,choi} or a multimode fiber~\cite{Cizmar2012,salma}. However, a transmission configuration is not adapted to non-invasive and/or in-vivo imaging of biological media. An epi-detection  geometry should thus be considered~\cite{Alexandrov}. During the last few years, the reflection matrix $\mathbf{R}$ had been investigated to perform selective focusing/ detection~\cite{popoff2,badon2016smart} or energy delivery~\cite{choi2,Jeong2018} through strongly scattering media. With regards to the specific purpose of imaging, the matrix approach has been recently used to implement AO tools in post-processing. The single scattering component of the reflected wave-field through biological tissues has been enhanced in depth by compensating for high-order aberrations~\cite{kang2015imaging,kang2017high}.

In this paper, we propose to go beyond a matrix approach of AO by introducing a novel operator: the so-called distortion matrix $\mathbf{D}$. Unlike previous works that investigated $\mathbf{R}$ either in the focal plane ~\cite{badon2016smart} or the pupil plane~\cite{popoff2,kang2015imaging,kang2017high}, we here consider the medium response between those dual bases~\cite{robert,robert2}. Unlike $\mathbf{R}$, the $\mathbf{D}$-matrix does not consider the reflected wave-field as a building block but its deviation from an ideal wave-front that would be obtained in absence of aberrations and without multiple scattering. This operation may seem trivial but it dramatically highlights the input/output correlations of the wave-field. While the canonical reflection matrix exhibits a random feature in a turbid medium, the distortion matrix displays strong field-field correlations over each IP. Thanks to this new operator, some relevant results of information theory can thus be fruitfully applied to optical imaging. \coraa{A singular value decomposition (SVD) of $\mathbf{D}$ \alex{allows a partition} of the field-of-illumination (FOI) into \cora{orthogonal} isoplanatic modes (IMs) and  extract the associated wave-front distortion in the pupil plane.} The Shannon entropy $\mathcal{H}$ of the singular values allows one to define the effective rank of the imaging problem.  A combination of the $\mathcal{H}$ first eigenstates yields an image of the focal plane with an excellent contrast and a diffraction-limited resolution as if the medium ahead was made perfectly transparent.

%A time-reversal analysis of $\mathbf{D}$ yields a one-to-one association between each isoplanatic mode and the corresponding aberrated wave-front in the pupil plane.
%The $\mathbf{D}$-matrix is defined and its correlations are investigated both in the pupil and focal planes. A physical picture is provided for the SVD of $\mathbf{D}$. It can be seen as a iterative time-reversal process on a virtual reflector which results from the incoherent sum of the incident focal pots. 
Several experiments with an increasing order of complexity are presented to demonstrate the benefits of the $\mathbf{D}$-matrix for optical imaging in turbid media. For sake of simplicity, the first experiment involves the imaging of a single IP through a thick layer of biological tissues. This configuration allows us to lay down the $\mathbf{D}$-matrix concept and highlight the physics behind it. Then, a second proof-of-concept experiment considers a thin but strong aberrating layer introduced between the microscope objective and a resolution target. This imaging configuration involves a spatially-varying aberration across the FOI (\textit{i.e} several IPs). At last, we describe an imaging experiment through a \cora{turbid} non-human primate cornea that induces high-order aberrations (including forward multiple scattering) and a strong diffuse multiple scattering background. The $\mathbf{D}$-matrix decomposes the imaging problem into a set of IMs whose degree of complexity increases with their rank (\textit{i.e.} smaller spatial extent in the focal plane and higher phase distortion in the pupil plane). This last experiment demonstrates the ability of our matrix approach to discriminate between forward multiple scattering paths, that can be taken advantage of for imaging, and the diffuse background, that shall be removed from the final image. % While the first contribution can be taken advantage of for imaging the focal plane, the second contribution never reaches it and should be removed from the final image. The $\mathbf{D}$-matrix concept allows to separate both contributions. The forward multiple scattering contribution exhibits field-field correlations due to the memory effect. It thus emerges along the signal subspace (highest singular values). On the contrary, the diffuse multiple scattering background is fully random and lies along the noise subspace of $\mathbf{D}$ the signal subspace of $\mathbf{D}$  that lies along its noise subspace

\begin{figure*}[htbp]
\center
\includegraphics[width=0.9\textwidth]{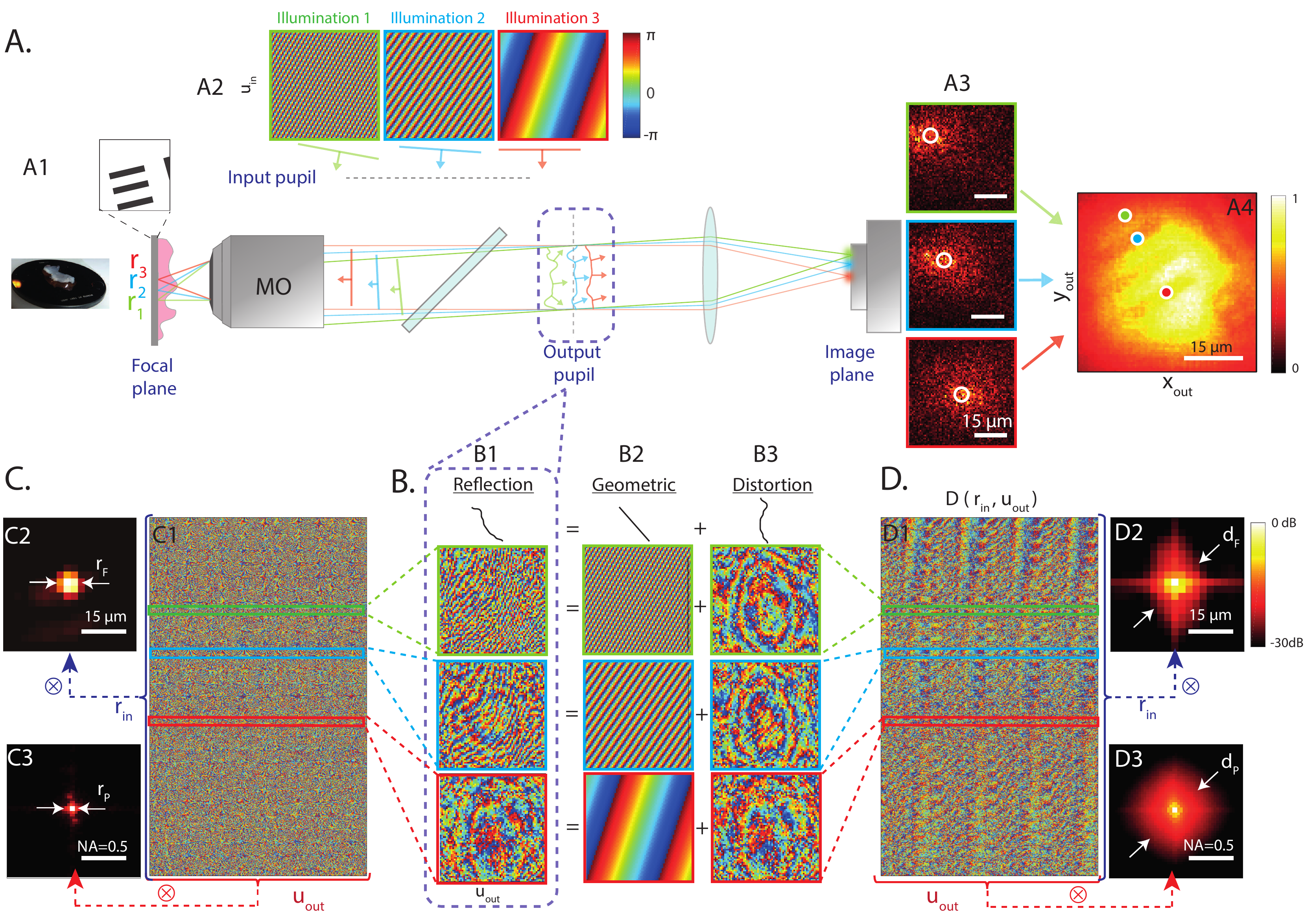}    
\caption{ \textbf{Principle of the distortion matrix approach.} (\textbf{A}) A resolution target (USAF 1951) is positioned underneath a 800-$\mu$m-thick sample of rat intestine ($\textbf{A1}$). In scanning microscopy, raster scanning in the focal plane is obtained using a set of plane wave illuminations in the input pupil ($\textbf{A2}$). In presence of \cora{sample-induced} aberrations, the detected intensity will exhibit a much larger extent compared to the ideal PSF (\textbf{A3}). The resulting full-field image displays a low contrast and a reduced resolution (\textbf{A4}). (\textbf{B}), In the output pupil plane, the phase of the reflected wave-field ($\textbf{B1}$) can be split into a diffraction (\textbf{B2}) and a distortion (\textbf{B3})  term. \cora{(\textbf{C},\textbf{D}) The reflected distorted wave-fields can be stored along column vectors to form the reflection and distortion matrices, $\mathbf{R}$ and $\mathbf{D}$, respectively. The phase of $\mathbf{R}$ and $\mathbf{D}$ is displayed in (\textbf{C1}) and (\textbf{D1}), respectively. The auto-correlations of the \coraa{complex} reflected/distorted wave-fields are computed in the focal (\textbf{C2}/\textbf{D2}\cora {, see Section~S2}) and in the pupil (\textbf{C3}/\textbf{D3}\cora{, see Section~S1}) planes, both in dB. All the data shown here are extracted from the rat intestine imaging experiment. \cora{Photo Credit: Amaury Badon, CNRS.}}
}
\label{fig1}
\end{figure*}

\section*{RESULTS}
\subsection*{Time-gated reflection matrix}

The $\mathbf{D}$-matrix concept first relies on the measurement of the time-gated reflection matrix $\mathbf{R}$ from the scattering sample. Until now, optical transmission/reflection matrices have always been investigated either in the $\mathbf{k}$-space (plane-wave basis) \cite{popoff2010measuring,kang2015imaging} or in the real space (focused basis) \cite{badon2016smart}. Here the $\mathbf{R}$-matrix will be defined between those dual bases. \cora{ This choice is dictated by our will to go beyond the study of restricted isoplanatic fields of view and tackle space-variant aberrations. Indeed, waves produced by nearby points inside a complex medium can generate highly correlated, but
tilted, random speckle patterns in the far-field~\cite{Osnabrugge2017}. In a focused basis, this corresponds to a spatially invariant point spread function (PSF) over an area called the isoplanatic patch. As we will see, only a dual-basis matrix can highlight these angular correlations that persist over a restricted spatial domain in the focal plane.}

The experimental set-up has already been described in a previous work ~\cite{badon2016smart} and is displayed in Fig.~S1. The experimental procedure is detailed in the Methods section. \coraa{In a few words}, the sample is illuminated through a microscope objective (MO) by a set of focused waves (input focusing basis) (see Fig.~\ref{fig1}A). For each illumination, the amplitude and phase of the reflected wave-field is recorded by phase-shifting interferometry on a CCD camera placed in the pupil plane (output pupil basis). A coherent time gating is also applied in order to select ballistic and snake photons while eliminating a (large) part of the diffuse photons. A set of time gated reflection coefficients, $R(\mathbf{u_\textrm{out}},\mathbf{r_\textrm{in}})$, is finally measured between each virtual source in the focal plane identified by the vectors $\mathbf{r_\textrm{in}}$ at the input and each point of the pupil plane $\mathbf{u_\textrm{out}}$ at the output. These coefficients form the reflection matrix $\mathbf{R}$ (see Fig.~\ref{fig1}D). 

The first imaging problem we consider in this paper deals with an experiment through biological tissues (see Fig.~\ref{fig1}A). A positive U.S. Air Force (USAF) 1951 resolution target placed behind an 800-$\mu$m-thick of rat intestinal tissue is imaged through an immersion objective [40$\times$, NA (numerical aperture), 0.8; Nikon]. The rat intestinal tissue displays  a refractive index $n\sim1.4$, a scattering mean free path $\ell_s$ of the order of 150 $\mu$m and an anisotropy factor $g\simeq 0.9$ \cite{Jacques}. The reflection matrix $\mathbf{R}$ is measured over \cora{a FOI $\Omega\times\Omega = 41 \times 41$ $\mu$m$^2$} with $N_\textrm{in}=$729 input wave-fronts, a spatial sampling $\delta r_\textrm{in}=1.6$ $\mu$m and an input pupil aperture \cora{$\mathcal{D}_\textrm{in}\times \mathcal{D}_\textrm{in}=1.7\times1.7$ mm$^2$}. \coraa{This reduced pupil diameter corresponds to the size of the illumination beam (see Fig.~S2)}. At the output, the wave-field is recorded over a pupil size of \cora{$\mathcal{D}_\textrm{out}\times \mathcal{D}_\textrm{out}=4.5 \times 4.5$} mm$^2$ with $N_\textrm{out}=6084$ pixels and a spatial sampling \cora{$\delta u_\textrm{out}=68$ $\mu$m. \cora{The corresponding field-of-view is $60 \times 60$ $\mu$m$^2$}. This experimental configuration corresponds to an imaging condition for which time gating guarantees that the reflection matrix contains a fraction of ballistic or snake photons reflected by the resolution target (see Fig. S3). However, aberrations are so intense that the full-field image of the resolution target is dominated by the diffuse multiple scattering background (see Fig.~\ref{fig1}A4).}

Figure~\ref{fig1}B1 displays examples of reflected wave-fields for several virtual sources $\mathbf{r_\textrm{in}}$. Each wave-field is stored along a column vector and forms the reflection matrix $\mathbf{R} = [R(\mathbf{u_\textrm{out}},\mathbf{r_\textrm{in}})]$. $\mathbf{R}$ exhibits a 4D-structure but is concatenated both at the input and output to be displayed in 2D (see Fig.~S4). The phase of $\mathbf{R}$ is displayed in Fig.~\ref{fig1}C1. Its spatial and angular correlations in the focal and pupil planes are displayed in Figs.~\ref{fig1}C2 and C3, respectively. As it could be conjectured from the column vectors displayed in Fig.~\ref{fig1}B1, the matrix $\mathbf{R}$ only displays short-range correlations. This is quite surprising as the object to be imaged is deterministic and contained in a single IP. 
To understand this seemingly randomness of $\mathbf{R}$ and reveal its hidden correlations, we now investigate its theoretical expression. The reflection matrix can be expressed as follows (see Fig.~S5):
\begin{equation}
\mathbf{R} = \cora{\mathbf{T}} \times \mathbf{\Gamma} \times \mathbf{H_\textrm{in}}
\label{int_abe}
\end{equation}
or, in terms of matrix coefficients,
\begin{equation}
R(\mathbf{u_\textrm{out}},\mathbf{r_\textrm{in}})  =  \int \cora{T}(\mathbf{u_\textrm{out}},\mathbf{r})  \gamma(\mathbf{r}) H_\textrm{in}(\mathbf{r,r_\textrm{in}}) d\mathbf{r} 
\label{int_abe2}
\end{equation}
$\mathbf{H_\textrm{in}}=[H_\textrm{in}(\mathbf{r},\mathbf{r_\textrm{in}})]$ is
the input focusing matrix. Its columns are none other than the input focal spots centered around each focusing point $\mathbf{r_\textrm{in}}$ (see Fig.~S5). Under a single scattering assumption, $\mathbf{\Gamma}$ is a diagonal matrix whose elements $\gamma(\mathbf{r})$ \cora{map} the reflectivity of the \cora{object in the focal planes. This object is here assumed to cover the whole FOI.}
%The illumination/reflectivity product $S(\mathbf{r},\mathbf{r_\textrm{in}})=\rho(\mathbf{r}) H_\textrm{in}(\mathbf{r,r_\textrm{in}})$ accounts for the amplitude distribution of each virtual source $\mathbf{r_\textrm{in}}$. 
\cora{$\mathbf{T}$} is the transmission matrix between the focal and pupil planes (see Fig.~S5). Its elements $T(\mathbf{u_\textrm{out}},\mathbf{r})$ describe the propagation of the wave-field from a point $\mathbf{r}$ in the MO focal plane to a detector $\mathbf{u_\textrm{out}}$ in the output pupil plane. \cora{Theoretically, the correlation length $r_P$ of the reflected wave-field in the pupil plane scales as $\lambda f/{\Omega}$ (see Section~S1) while its correlation length $r_F$ in the focal plane is dictated by the coherence length of the input focal spots, that is to say the input diffraction limit, $\delta^\textrm{0}_\textrm{in}\sim \lambda f/{\mathcal{D}_\textrm{in}}$, in a strong aberration regime (see Section~S2). This accounts for the spatial incoherence exhibited by $\mathbf{R}$ both at its input (Fig.~\ref{fig1}C2) and output (Fig.~\ref{fig1}C3), respectively. In the next section, we show how to reveal the hidden correlations in $\mathbf{R}$ in order to, subsequently, extract the transmission matrix $\mathbf{T}$}.

\subsection*{Principle of the distortion matrix}

The holy grail for imaging is \cora{indeed} to have access to this transmission matrix \cora{$\mathbf{T}$}. Its inversion or pseudo-inversion would actually allow to reconstruct a reliable 3D image of the scattering medium, thereby overcoming aberration and multiple scattering effects generated by the medium itself. However, in most imaging configurations, the \textit{true} transmission matrix \cora{$\mathbf{T}$} is not accessible as it would require an invasive measurement. The common assumption in wave imaging, is thus to consider an homogeneous medium model. The free space transmission matrix \cora{$\mathbf{T_0}$} should then be considered. Its elements \cora{$  T_0(\mathbf{u_\textrm{out}},\mathbf{r})$} are simply given by
\begin{equation}
\label{G0}
    \cora{T_0}(\mathbf{u_\textrm{out}},\mathbf{r_\textrm{in}}) =\frac{1}{j \lambda f}\exp \left [ j \frac{2 \pi}{\lambda f} \mathbf{u_\textrm{out}} .\mathbf{r}\right ]
\end{equation}
where $f$ is the MO's focal length and $\lambda$ the central wavelength.

\begin{figure}[h!]
\center
\includegraphics[width=\textwidth]{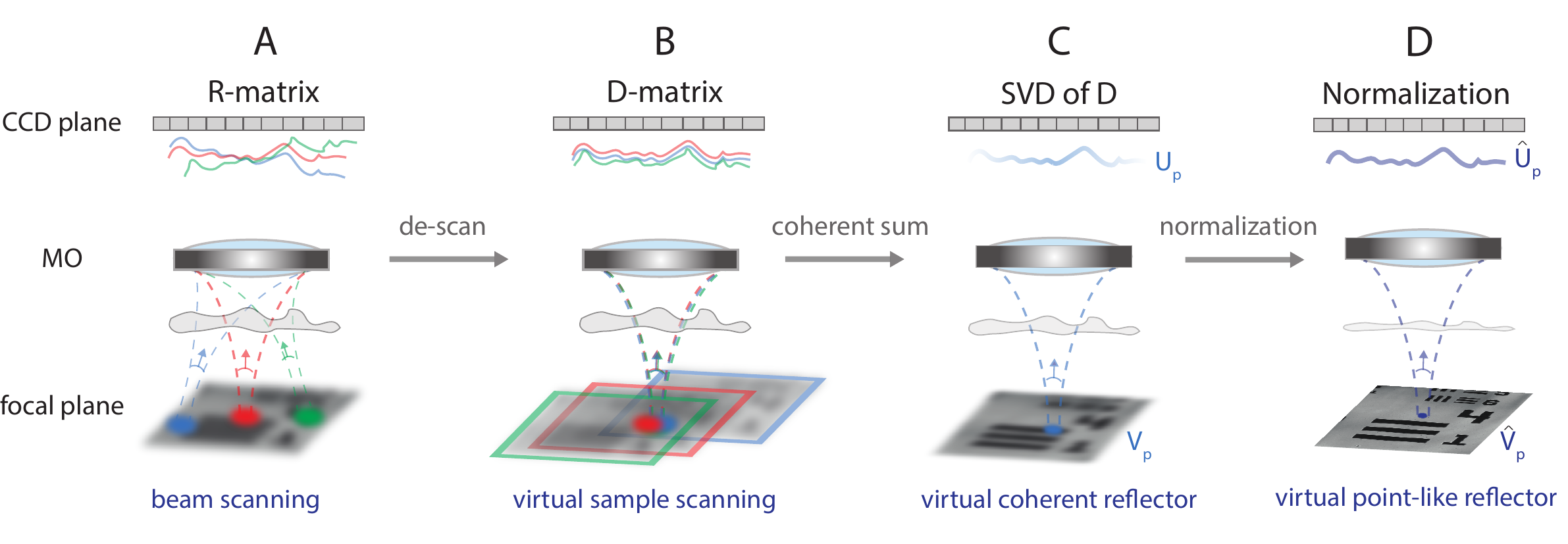}
\caption{\label{fig3}\cora{\textbf{Extracting the aberration transmittance from the distortion matrix $\mathbf{D}$}. ($\textbf{A}$) The recording of the $\mathbf{R}$-matrix consists in scanning the objects with a moving input focusing beam. ($\textbf{B}$) The removal of the geometric component in each reflected wave-front [Eq.\ref{D}] amounts to recenter each incident focal spot at the origin. The $\mathbf{D}$-matrix is equivalent to the reflection matrix for a moving object. ($\textbf{C}$) The SVD of $\mathbf{D}$ leads to a coherent sum of the distorted wave-fronts in the pupil plane. A coherent reflector is virtually synthesized in the focal plane and the corresponding wave-front emerges along the output singular vector $\mathbf{U_1}$. The corresponding image of the object is provided by the first input singular vector $\mathbf{V_1}$ but its resolution is dictated by the width $\delta_\textrm{in}$ of the input focusing beam. ($\textbf{C}$) A normalization of $\mathbf{U_1}$ in the pupil plane makes the virtual scatterer point-like. The corresponding input singular vector $\mathbf{\hat{V}_1}$ yields a diffraction-limited image of the object in the focal plane.}}
\end{figure}

In this work, we will use \cora{$\mathbf{T_0}$} as a reference matrix. The columns of \cora{$\mathbf{T_0}$} are actually the reflected wave-fields that would be obtained in an ideal case, \textit{i.e} without aberrations. \coraa{In the Fourier space, the phase of the complex wave-field, or wave-front, is particularly adequate to study the effect of aberrations.} Figure~\ref{fig1}B compares few examples of reflected wave-fronts (columns of $\mathbf{R}$, see Fig.~\ref{fig1}B1) with the corresponding ideal wave-fronts (columns of \cora{$\mathbf{T_0}$}, see Fig.~\ref{fig1}B2). While the latter ones display plane wave fringes whose orientation and spatial frequency is related to the position $\mathbf{r_\textrm{in}}$ of the input focusing point, the recorded wave-fronts consist in a stack of this geometrical component with a distorted phase component induced by the biological tissues. The key idea of this paper is to isolate the latter contribution by subtracting the recorded wave-front by its ideal counterpart. Mathematically, this operation can be expressed as a Hadamard product between $\mathbf{R}$ and \cora{$\mathbf{T_0^*}$} (where $*$ stand for phase conjugate),
\begin{equation}
D  =  \mathbf{R} \circ  \cora{\mathbf{T_0^*}}
\label{D}
\end{equation}   
which, in term of matrix coefficients, can be written as
\begin{equation}
D(\mathbf{u_\textrm{out}},\mathbf{r_\textrm{in}})  =  R(\mathbf{u_\textrm{out}},\mathbf{r_\textrm{in}}) \times \cora{T_0^*}(\mathbf{u_\textrm{out}},\mathbf{r_\textrm{in}})
\label{D2}
\end{equation}   
The matrix $\mathbf{D}=[D(\mathbf{u_\textrm{out}},\mathbf{r_\textrm{in}})]$ is the so-called distortion matrix. Removing the geometrical component of the reflected wave-field in the pupil plane as done in Eq.\ref{D} amounts to \cora{a change of reference frame. While the original  reflection matrix is recorded in the object's frame (static object scanned by the input focusing beam, see Fig.~\ref{fig3}A), the $\mathbf{D}$-matrix is a reflection matrix in the frame of the input focusing beam (moving object illuminated by a static beam, see Fig.~\ref{fig3}B).} Physically, this corresponds to a descan of the reflected light as performed in confocal microscopy.%$\mathbf{D}$ is thus a reflection matrix but with different realizations of virtual sources all located at the origin. 
 %Removing the geometrical component of the reflected wave-field in the pupil plane as done in Eq.\ref{D} amounts to shift each virtual source on the optical axis in the focal plane. Physically, this corresponds to a descan of the reflected light as performed in confocal microscopy. $\mathbf{D}$ is thus a reflection matrix but with different realizations of virtual sources all located at the origin (see Fig.~\ref{S4}). 

The $\mathbf{D}$-matrix deduced from $\mathbf{R}$ is displayed in Fig.~\ref{fig1}D1. Compared to $\mathbf{R}$ (Fig.~\ref{fig1}C1), \cora{it exhibits long-range correlations both} in the pupil (Fig.~\ref{fig1}D3) and focal (Fig.~\ref{fig1}D2) planes, respectively. 
\cora{On the one hand, by virtue of the van Cittert Zernike theorem~\cite{born}, the coherence length $d_P$ of the distorted wave-field in the pupil plane scales as $\lambda f/\delta_\textrm{in}$, with $\delta_\textrm{in}$ being the spatial extension of the incoherent input focal spot $|H_\textrm{in}|^2$ (see Section~S2). On the other hand, its correlation length $d_F$ in the focal plane corresponds to the size $\ell_c$ of the isoplanatic patch (see Section~S2).} This is illustrated by examples of distorted wave-fields displayed in Fig.~\ref{fig1}B3. While the original reflected wave-fronts did not display any similarity, the distorted component displays similar Fresnel rings whatever the focusing point $\mathbf{r_\textrm{in}}$. The $\mathbf{D}$-matrix thus reveals input/output correlations of the wave-field that were originally completely hidden in the original $\mathbf{R}$-matrix (Fig.~\ref{fig1}C). 

\subsection*{\coraa{Singular value decomposition of the distortion matrix}}

\coraa{The next step is to extract and take advantage of those field-field correlations for imaging. To that aim, a singular value decomposition (SVD) \cora{of the distortion matrix} is performed. It consists in writing $\mathbf{D}$ as follows}
\begin{equation}
\label{svd}
\mathbf{D}=\mathbf{U \Sigma  V^{\dag}}
\end{equation}
or, in terms of matrix coefficients,
\begin{equation}
\label{svd2}
D(\mathbf{u_\textrm{out}},\mathbf{r_\textrm{in}})=\sum_{p=1}^{N_\textrm{in}} \sigma_p U_p(\mathbf{u_\textrm{out}}) V^*_p(\mathbf{r_\textrm{in}}).
\end{equation}
$\mathbf{\Sigma}$ is a diagonal matrix containing the real positive singular values $\sigma_i$ in a decreasing order $\sigma_1>\sigma_2>..>\sigma_{N_\textrm{in}}$. $\mathbf{U}$ and $\mathbf{V}$ are unitary matrices whose columns, $\mathbf{U_p}=[U_p(\mathbf{u_\textrm{out}})]$ and $\mathbf{V_p}=[V_p(\mathbf{r_\textrm{in}})]$, correspond to the output and input singular vectors, respectively. \alex{The symbol $\dag$ stands for transpose conjugate.} \cora{Mathematically, the SVD extracts a signal subspace associated with the largest singular values and characterized by an important correlation between its lines and/or columns. In the $\mathbf{D}$-matrix, these correlations are induced by the isoplanicity of the input PSF $H_\textrm{in}$. The single scattering and forward multiple scattering contributions are expected to lie along the signal subspace since they exhibit a spatial invariance over each isoplanatic patch~\cite{Mertz15}. On the contrary, the diffuse photons induced by the scattering layer ahead of the focal plane give rise to a fully incoherent wave-field that will be equally distributed over all the eigenstates of $\mathbf{D}$~\cite{badon2017multiple}. Hence, the pollution of the signal subspace by the multiple scattering noise \alex{scales as the inverse of the number of independent input focusing points mapping each isoplanatic patch, $(\ell_c/\delta_\textrm{in}^\textrm{0})^2$. A large isoplanatic patch enables the SVD to drastically decrease the multiple-to-single scattering ratio.}}

\cora{To know which of the input or output correlations will dictate the SVD of $\mathbf{D}$, relevant parameters are the numbers of independent speckle grains, $M_D$ and $N_D$, exhibited by $\mathbf{D}$ at its input and output, respectively. The correlation degree of the distorted wave-field in each plane is actually inversely proportional to the corresponding number of independent speckle grains. In the focal plane, $M_D$ is given by the squared ratio between the FOI $\Omega$ and the coherence length $d_F$ of the distorted wave-field in the focal plane 
\begin{equation}
    M_D=\left ( \Omega/d_F \right )^2.
\end{equation}
$d_F$ is the minimum between the isoplanatic length $\ell_c$ and the characteristic fluctuation length $\ell_\gamma$ of the object's reflectivity (see Section~S2). In the pupil plane, the number $N_D$ of independent speckle grains scales as  (see Section~S1)
\begin{equation}
\label{ND0}
    N_D=\left ( \delta_\textrm{in}/\delta^\textrm{0}_\textrm{out} \right )^2,
\end{equation}
where $\delta^\textrm{0}_\textrm{out}$ is the diffraction-limit resolution at the output (Eq.~\ref{resolution}). The domination of input correlations implies the following condition:
\begin{equation}
\label{strong_ab}
M_D<N_D.
\end{equation}
If $\ell_\gamma > \ell_c$, the last equation can be translated as follows: The number $M_D=(\Omega/\ell_c)^2$ of IPs supported by the FOI should be smaller than the number $N_D$ of resolution cells that map each input focusing beam (Eq.~\ref{ND0}). As we will see, this strong aberration condition is fulfilled in the experiments presented in this work.} 

\cora{When input correlations dominate, the effective rank of the signal subspace then corresponds to the number of independent spatial modes required to map the distorted wave-field in the focal plane, \textit{i.e} the number $M_D$ of IPs. As shown in Section~S3, the SVD decomposes the \cora{FOI} onto a set of orthonormal IMs defined by the input singular vectors $\mathbf{V_p}$. The corresponding output singular vectors $\mathbf{U_p}$ yield the associated aberration phase laws in the pupil plane. Their coherent combination can then lead to the retrieval of the transmission matrix $\mathbf{T}$.} 
%For instance, in the case of a mirror object and a spatially constant aberration, the distortion matrix will be constant both in amplitude (reflectivity of the specimen) and in phase (distortion of the wavefront). In this case, the matrix is of rank 1 and a single subspace is enough to describe the distortion matrix. On the contrary, if the aberrated wavefront is now different for every location in the FOV, the rank of the matrix will be equal to the number of resolution cells contained in the FOV.}

In the next sections, we will check all these promising properties of $\mathbf{D}$ experimentally, and see how we can take advantage of it for deep imaging.
 
\begin{figure}[htbp]
\center
\includegraphics[width=0.8\textwidth]{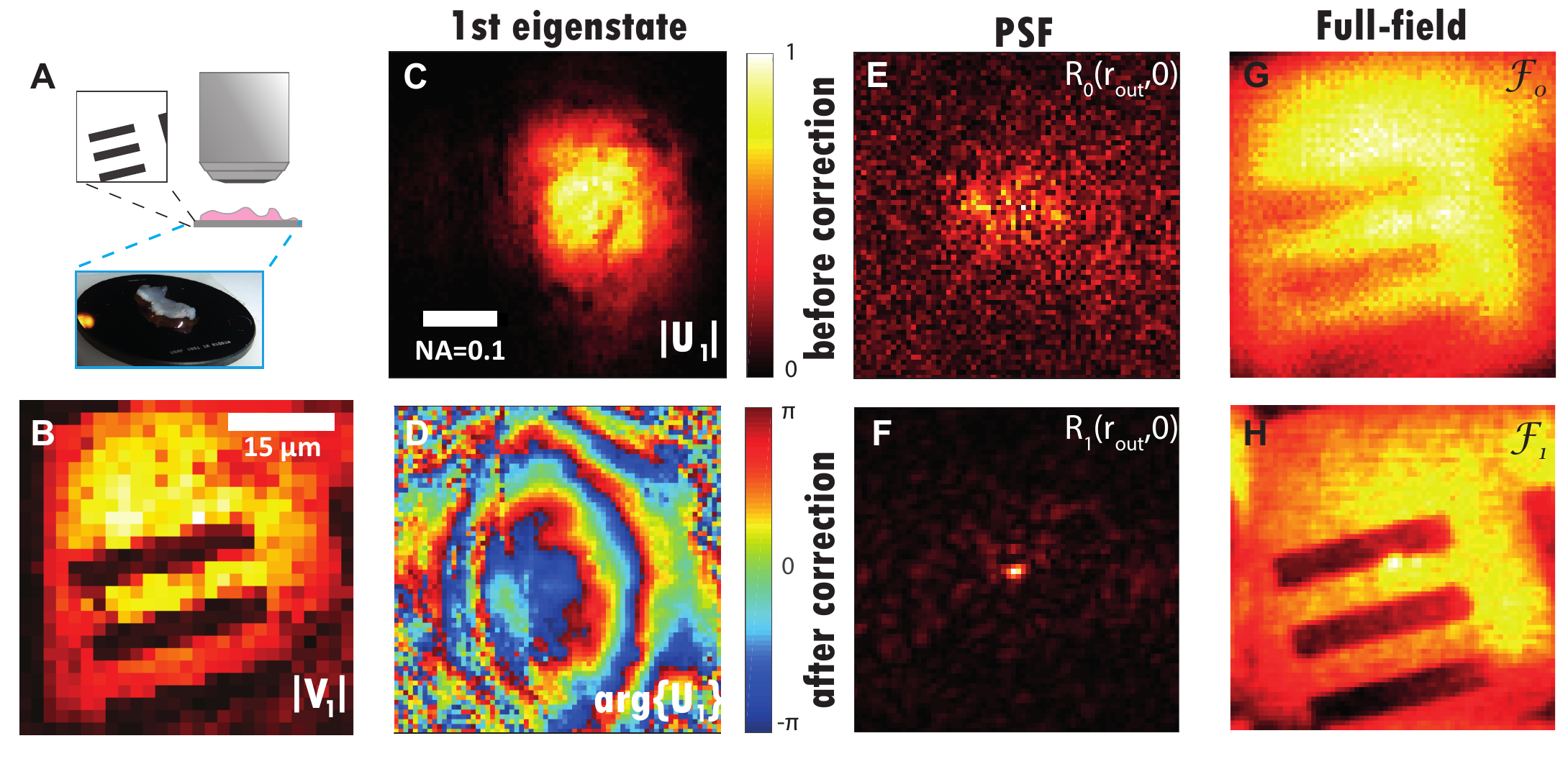}
\caption{\textbf{Imaging through a thick layer of rat intestinal tissue}. \cora{(\textbf{A}) Experimental configuration. (\textbf{B},\textbf{C}) Modulus of the first input singular vector $\mathbf{V_1}$ of $\mathbf{D}$ in the focal plane. (\textbf{D}) Modulus and phase of the first output singular vector $\mathbf{U_1}$ in the pupil plane.} (\textbf{E}) Example of PSF deduced from the central column ($\mathbf{r_\textrm{in}}=\mathbf{0}$) of the raw focused matrix $\mathbf{R_0}$. (\textbf{F})  Corresponding corrected PSF deduced from the central column of the focused matrix $\mathbf{R_1}$ (Eq.~\ref{PSF}). (\textbf{G},\textbf{H}) Comparison of the full-field images $\mathcal{F}_0$ and $\mathcal{F}_1$ (Eq.~\ref{F}) before and after aberration correction. \cora{Photo Credit: Amaury Badon, CNRS.}
\label{fig2} }
\end{figure}

\subsection*{Imaging over a single isoplanatic patch}

\cora{The reflection and distortion matrices corresponding to the imaging experiment through a thick layer of rat intestine are shown in Figs.~\ref{fig1}C1 and D1, respectively. The long-range spatial correlations exhibited by $\mathbf{D}$ (Fig.\ref{fig1}D2) seem to indicate that the isoplanatic hypothesis is close to being fulfilled in this experiment. The SVD of $\mathbf{D}$ confirms this intuition by exhibiting a predominant eigenstate. The corresponding singular vectors $\mathbf{V_1}$ and $\mathbf{U_1}$ are displayed in Fig.~\ref{fig2}.} The modulus of $\mathbf{V_1}$ displays a contrasted image of the resolution target (Fig.~\ref{fig2}B) \cora{but its resolution is limited by the low spatial sampling of the illumination scheme. The output singular vector $\mathbf{U_1}$ corresponds to the wave-front induced by a virtual coherent reflector of scattering distribution $|H_\textrm{in} (\mathbf{r}-\rin)|^2$, hence located on the optical axis in the focal plane (see Fig.~\ref{fig3}C). \cora{This virtual scatterer results from a coherent summation of the de-scanned input focal spots through the SVD process (see Section~S3).} Its phase is made of Fresnel rings mainly induced by the irregular surface of the sample and its index mismatch with the surrounding fluid (Fig.~\ref{fig2}D). \cora{Its finite support is explained by the finite size $\delta_\textrm{in}$ of the coherent reflector (Fig.~\ref{fig2}C). To make this virtual scatterer point-like} and retrieve a diffraction-limited image (Fig.~\ref{fig3}D), a normalized vector $\mathbf{\tilde{U}_1}$ should be considered, such that $\tilde{U}_1(\uout)= U_1(\uout)/ |U_1(\uout)|$. $\mathbf{\tilde{U}_1}$ can be used to build an estimator $\mathbf{\hat{T}}$} of the transmission matrix between the pupil and focal planes, such that its coefficients read
\begin{equation}
\label{transmit}
    \cora{\hat{T}_p(\mathbf{u_\textrm{out}},\mathbf{r_\textrm{in}})=  \tilde{U}_p(\uout)  {T_0}(\mathbf{u_\textrm{out}},\mathbf{r_\textrm{in}})}
    \end{equation}
    with $p=1$ in the present case.
This estimator can be used to project the recorded matrix $\mathbf{R}$ in the focal basis both at input and output, such that
\begin{equation}
\label{PSF}
    \mathbf{R_p}= \mathbf{\hat{\cora{T}}_p^{\dag}}\mathbf{R} 
    \end{equation}
The coefficients $R_1(\mathbf{r_\textrm{out}},\mathbf{r_\textrm{in}})$ are the impulse responses between each input focusing point $\mathbf{r_\textrm{in}}$ and each output imaging point $\mathbf{r_\textrm{out}}$. In other words, once reshaped in 2D, each column of $ \mathbf{R_1}$ yields the PSF of the imaging system at the input focusing point $\mathbf{r_\textrm{in}}$. The PSF for an input focusing point on the optical axis ($\mathbf{r_\textrm{in}}=\mathbf{0}$) is displayed in Fig.~\ref{fig2}\cora{F}. For sake of comparison, the corresponding initial focal spot is displayed in Fig.~\ref{fig2}\cora{E}. The latter one is extracted from the focused matrix $\mathbf{R_0}$ deduced from $\mathbf{R}$ using \cora{$\mathbf{T_0}$}: $\mathbf{R_0}=\mathbf{\cora{T_0^{\dag}}}\mathbf{R}$. While the initial PSF exhibits a random speckle pattern (Fig.~\ref{fig2}\cora{E}), the PSF after correction shows a nearly diffraction-limited focal spot with almost all the energy concentrated in the vicinity of the incident focusing point (Fig.~\ref{fig2}\cora{F}). The \cora{apparent width of this PSF yields an estimation of the local output resolution $\delta_\textrm{out}$ at $\mathbf{r_\textrm{in}}$. Here, $\delta_\textrm{out}$ goes from \cora{20} $\mu$m on the raw data (Fig.~\ref{fig2}\cora{E}) to \cora{1.2} $\mu$m after the matrix correction (Fig.~\ref{fig2}\cora{F}).} This value should be compared to the diffraction-limited resolution
\begin{equation}
\label{resolution}
\cora{\delta^\textrm{0}_\textrm{out}=\frac{\lambda}{2 \mathrm{NA}_\textrm{out}}},
\end{equation}
\cora{with $\mathrm{NA}_\textrm{out}=\mathcal{D}_\textrm{out}/(2f)=0.45$ being the output numerical aperture.} The numerical application of this formula yields \cora{$\delta^\textrm{0}_\textrm{out} \simeq 0.9 $ $\mu$m} in our experimental configuration. The mismatch between \cora{$\delta_\textrm{out}$} and \cora{$\delta^\textrm{0}_\textrm{out}$} comes from the noisy aspect of $\mathbf{U_1}$ at large spatial frequencies (see Fig.~\ref{fig2}D), which prevents from an efficient aberration compensation over the whole \cora{numerical aperture}.

If the spatial sampling was equivalent at input and output, a confocal image could be deduced from the diagonal elements ($\mathbf{r_\textrm{in}}=\mathbf{r_\textrm{out}}$) of $\mathbf{R_0}$ and $\mathbf{R_1}$~\cite{badon2016smart}. Here, as a sparse illumination scheme has been employed \cora{($\delta r_\textrm{in}>\delta^\textrm{0}_\textrm{out}$)}, a full-field image is considered and obtained by summing $\mathbf{R}$ over its input elements:
\begin{equation}
\label{F}
    \mathcal{F}_p(\mathbf{r_\textrm{out}})=\sum_{\mathbf{r_\textrm{in}}} |R_p(\mathbf{r_\textrm{out}},\mathbf{r_\textrm{in}})|
    \end{equation}
with $p=0$ or $1$ here. The corresponding images $\mathcal{F}_0$ and $\mathcal{F}_1$ are displayed in Figs.~\ref{fig2}\cora{G and H}, respectively. While the patterns of the resolution target are hardly visible on the raw image, the $\mathbf{D}$-matrix approach provides a highly contrasted image of the target. To quantify this gain in image quality, the Strehl ratio is a relevant parameter~\cite{mahajan1982strehl}. It is defined as the ratio of the PSF peak intensity with and without aberration. Equivalently, it can also be defined in the pupil plane as the squared magnitude of the mean aberration phase factor. Its initial value $\mathcal{S}_0$ can thus be directly derived from the $\mathbf{D}$-matrix coefficients:
\begin{equation}
\label{S0}
    \mathcal{S}_0= \left | \left \langle \exp \left (j \mbox{arg} \left \lbrace D (\mathbf{u_\textrm{out}},\mathbf{r_\textrm{in}}) V_1(\mathbf{r_\textrm{in}})\right \rbrace \right ) \right \rangle \right |^2
\end{equation}
where the symbol $\langle \cdots \rangle$ denotes an average over $\mathbf{u_\textrm{out}}$ and $\mathbf{r_\textrm{in}}$. In the present case, the original Strehl ratio is $\mathcal{S}_0=8\times 10^{-5}$. This experiment corresponds to imaging conditions far from being in the range of operation of conventional AO and explains why the patterns of the resolution target are so hardly visible on the raw image (Fig.~\ref{fig2}\cora{G}). The Strehl ratio $\mathcal{S}_1$ after the $\mathbf{U_1}$ correction can be directly extracted from the SVD of $\mathbf{D}$ (Eq.~\ref{svd2}):
\begin{equation}
\label{S}
    \mathcal{S}_1= \left | \left \langle   
    \exp \left (j  \mbox{arg} \left \lbrace U_1^*  (\mathbf{u_\textrm{out}})D (\mathbf{u_\textrm{out}},\mathbf{r_\textrm{in}}) V_1(\mathbf{r_\textrm{in}}) \right \rbrace \right ) \right \rangle \right |^2
\end{equation}
The $\mathbf{D}$-matrix correction leads to a Strehl ratio $\mathcal{S}_1=3\times 10^{-3}$. However, Eq.~\ref{S} gives the same weight to bright and dark areas of the resolution target in the focal and pupil planes. One possibility is to consider a weighted average instead of Eq.\ref{S} by the object reflectivity $|V_1(\mathbf{r_\textrm{in}})|^2$. This weighted Strehl ratio $\mathcal{S}'_1$ then reaches the value of $1.1 \times 10^{-2}$. \cora{Such a Strehl ratio value is relatively low but it should be kept in mind that the distortion matrix is associated with a PSF in reflection that convolves the transmit and receive PSFs.
Our measurement of the Strehl ratio is thus degraded by: (\textit{i}) the subsistence of input aberrations; (\textit{ii}) the presence of a diffuse multiple scattering background that acts here as an additive noise. Note, however, that the gain in terms of Strehl ratio is absolute; this is the relevant quantity to assess the benefit of our matrix approach.} This gain here is spectacular ($\mathcal{S}'_1/\mathcal{S}'_0\sim$140) and accounts for the satisfying image of the resolution target obtained after the $\mathbf{D}$-matrix correction (see Fig.~\ref{fig2}\cora{H}). \cora{Figure S3 shows how this drastic improvement of the Strehl ratio allows us to push back the imaging depth limit from 450 $\mu$m to almost 1 mm.}

This first experiment demonstrates the benefit of the $\mathbf{D}$-matrix in the simple case of a FOI containing a single IP. In the next section, the case of multiple IPs is tackled.

\subsection*{Imaging over multiple isoplanatic patches}
\begin{figure*}[htbp]
\center
\includegraphics[width=0.85\textwidth]{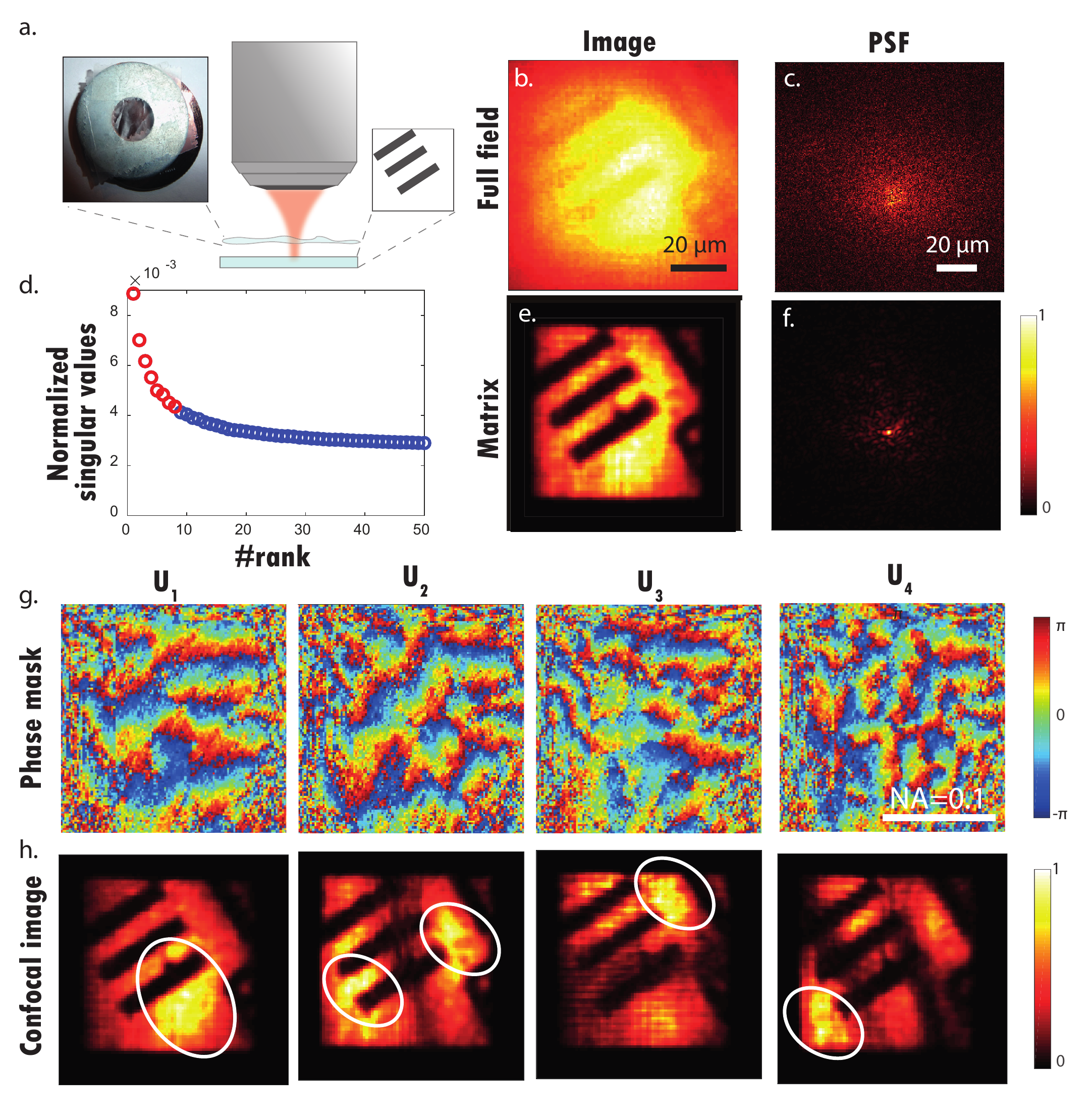}
\caption{\textbf{Matrix imaging over multiple isoplanatic patches}. (\textbf{A}) Schematic of the experiment. A resolution target (USAF 1951) is positioned at a distance $d = 1$  mm underneath a rough plastic film (see inset). (\textbf{B}) Original full-field image $\mathcal{F}_0$ (Eq.\ref{F}). (\textbf{C}) Example of PSF deduced from a column of the raw focused matrix $\mathbf{R_0}$. (\textbf{D}) Plot of the normalized singular values $\tilde{\sigma}_i$ of $\mathbf{D}$. The red circles correspond to the eight first singular values (signal subspace), while the noisy singular values are displayed in blue. (\textbf{E}) Matrix image constructed from the eight first eigenstates of $\mathbf{D}$ (Eq.~\ref{I2}). (\textbf{F}) Example of PSF deduced from a column of the corrected focused matrix $\mathbf{R_1}$. (\textbf{G}) Phase of the four first singular vectors $\mathbf{U_p}$.  (\textbf{H}) Confocal images deduced from the focused reflection matrices $\mathbf{R_p}$ [Eq.\ref{I}]. \cora{Photo Credit: Amaury Badon, CNRS.}}
\label{fig4}
\end{figure*}
The first element of the group 6 in the resolution target is now imaged through an aberrating layer consisting in a rough plastic sheet placed $d=$1 mm above the resolution target (USAF 1951) (see Fig.~\ref{fig4}A). 
The reflection matrix $\mathbf{R}$ is measured over a \cora{FOI} of $57 \times 57$ $\mu$m$^2$ with $N_\textrm{in}$=441 input wave-fronts, a spatial sampling $\delta r_\textrm{in}=2.85$ $\mu$m and an input pupil aperture $\mathcal{D}_\textrm{in}\times \mathcal{D}_\textrm{in}=1.3\times1.3$ mm$^2$. At the output, the wave-field is recorded over a pupil size of $\mathcal{D}_\textrm{out} \times \mathcal{D}_\textrm{out} =2 \times 2$ mm$^2$ with $N_\textrm{out}=12321$ pixels and a spatial sampling $\delta u_\textrm{out}$=18 $\mu$m. 

The full-field image $\mathcal{F}_0$ (Eq.~\ref{F}) and an example of PSF (Eq.~\ref{PSF}) are displayed in Figs.~\ref{fig4}A and B, respectively. The PSF is strongly degraded with a characteristic focal spot dimension $\delta_\textrm{out} \sim 45$ $\mu$m. This PSF dimension allows an estimation of the coherence length $\ell_c$ of the aberrating layer. \cora{Indeed, under a thin phase screen model~\cite{Mertz15}, the IP dimension $\ell_c$ coincides with the coherence length of the aberration transmittance. It turns out that the PSF width is inversely proportional to $\ell_c$ in this experiment: $\delta_\textrm{out}\sim \lambda d/\ell_c$. The IP size and the number of IPs supported by the FOI can be deduced from the PSF width $\delta$: $\ell_c \sim 18$ $\mu$m and \cora{$M_D \sim (\Omega/\ell_c)^2 \sim 10$.}}

A $\mathbf{D}$-matrix is deduced from $\mathbf{R}$ (Eq.~\ref{D}). Its analysis leads to the following estimation of the initial Strehl ratio: \cora{$\mathcal{S}^\prime_0= 1.6 \times 10^{-6}$ (Eq.~\ref{S0})}. This particularly strong aberration level accounts for the highly blurred aspect of the full-field image in Fig.~\ref{fig4}A. This experimental situation is thus particularly extreme, even almost \coraa{hopeless}, for a successful imaging of the resolution target. Yet the SVD of $\mathbf{D}$ will provide the solution.

Fig.~\ref{fig4}D displays the histogram of the normalized singular values $\tilde{\sigma}_i=\sigma_i/\sum_{j=1}^{N_\textrm{in}}\sigma_j$. If recorded data was not corrupted by experimental noise, the matrix would be of effective rank $M_D$. We could use all the eigenstates of $\mathbf{D}$ associated with non-zero singular values to retrieve an image of the object. In Fig.~\ref{fig4}D, only few singular values seem to emerge from the noise background. Hence, it is difficult to determine the number of eigenstates we need to consider to properly reconstruct an image of the object. This issue can be circumvented by computing the Shannon entropy $\mathcal{H}$ of the singular values~\cite{Campbell1960,Roberts1999}, such that
\begin{equation}
  \mathcal{H}(\tilde{\sigma}_i)= -\sum_{i=1}^{N_\textrm{in}} \tilde{\sigma}_i \log_2 \left ( \tilde{\sigma}_i \right ).
   \label{entropy}
\end{equation}
Shannon entropy delivers the maximally-noncommittal data set at a given signal-to-noise ratio, that is to say, the most information with the least artifact. The Shannon entropy can be used as an indicator of how many eigenstates are needed to build an adequate image of the object without being affected by experimental noise. 

The singular values of Fig.~\ref{fig4}D have an entropy $\mathcal{H}\simeq 8.4$. Hence, only the eight first singular states shall be considered. Fig.~\ref{fig4}G displays the phase of the four first singular vectors $\mathbf{U_i}$ in the pupil plane. They display phase distortions whose typical coherence length $u_c$ scales as $f \ell_c/d \sim 100 $ $\mu$m. The phase conjugation of these singular vectors should compensate for the detrimental effect of the aberrating layer in different parts of the FOI. A set of focused reflection matrices $\mathbf{R_p}$ can be deduced (Eq.\ref{PSF}). Fig.~\ref{fig4}f displays an example of corrected PSF extracted from a column of $\mathbf{R_1}$. Its comparison with the original PSF in Fig.~\ref{fig4}C shows how the phase conjugation of $\mathbf{U_1}$ allows one to compensate for the aberrations at this incident focusing point. On the one hand, the PSF width is narrowed by a factor 20, going from $\delta_\textrm{out} \sim 45$ $\mu$m to $2.25$ $\mu$m. The latter value should be compared with the diffraction-limited resolution \cora{$\delta^\textrm{0}_\textrm{out} \sim 2 $} $\mu$m (Eq.~\ref{resolution}) in our experimental conditions. The Strehl ratio is increased by a factor 2.2$\times 10^3$ \cora{to reach the final value $\mathcal{S}'_1=3.5 \times  10^{-3}$ (Eq.~\ref{S}). Again, this value of the Strehl ratio is probably underestimated because of input aberrations and multiple scattering.}

Confocal images can be deduced from the focused reflection matrices $\mathbf{R_p}$:
\begin{equation}
\label{I}
    \mathcal{I}_p(\mathbf{r_\textrm{out}})=\sum_{\mathbf{r_\textrm{in}}} |R_p(\mathbf{r_\textrm{out}},\mathbf{r_\textrm{in}})| e^{ - || \mathbf{r_\textrm{out}}-\mathbf{r_\textrm{in}} ||^2/2l_p^2}
    \end{equation}
where $l_p$ is the aperture of the numerical confocal pinhole~\cite{badon2016smart}. This finite aperture enables an average of the output image over neighbour incident focusing points in order to smooth out the sparse illuminations. Fig.~\ref{fig4}H displays the confocal images $ \mathcal{I}_p$ for $l_p=2$ $\mu$m. For a specular object such as a resolution target, the SVD has indeed the property of decomposing into a set of orthogonal IMs of \cora{spatial period $\ell_c$ (see Section~S3)}.  Their shape depends on the auto-correlation function of the aberrating phase screen. \coraa{A general trend is that the spatial frequency content of the eigenvectors increases with their rank. If this function presents an exponential or sinc model, the FOI will be spatially decomposed} into sinusoidal wave functions~\cite{Ghanem} analogous to optical fiber modes or to prolate spheroidal wave functions~\cite{robert2009}, respectively. Here, the autocorrelation function of the aberrating phase displays a Gaussian-like shape. The FOI is thus \coraa{spatially} mapped onto Hermite-Gaussian wave functions analogous to laser cavity modes~\cite{aubry2006}. 

\cora{The normalized pupil singular vectors $\mathbf{\tilde{U}_p}$ yield a set of orthogonal phase transmittances that map aberrations onto each isoplanatic mode. A coherent combination of these singular vectors should lead, in principle, to a satisfying estimator of the transmission matrix (see Section~S3)
\begin{equation}
    \mathbf{\hat{T}}=\sum_{p=1}^{\mathcal{H}(\tilde{\sigma}_i)} \mathbf{\tilde{U}_p} \circ \mathbf{T_0}.
\end{equation}
In practice,} a final image $\mathcal{I}$ of the resolution target can be obtained by summing the previous IMs $\mathcal{I}_p$:
\begin{equation}
\label{I2}
    \mathcal{I}(\mathbf{r_\textrm{out}})=\sum_{p=1}^{\mathcal{H}(\tilde{\sigma}_i)}  \mathcal{I}_p(\mathbf{r_\textrm{out}}).
    \end{equation}
The final result is displayed in Fig.~\ref{fig4}E. The comparison with the initial full-field image (Fig.~\ref{fig4}B) illustrates the benefit of the $\mathbf{D}$-matrix approach. Spatially-varying aberrations are overcome and a contrasted image of the resolution target is recovered over the whole FOI. This experiment demonstrates how the $\mathbf{D}$-matrix enables a decomposition of the FOI into several IMs and a mapping of each of them onto orthonormal distorted phase laws. However, this demonstration has been restricted to the case of a 2D aberrating phase layer. In the next section, we consider the case of a cornea with \cora{deteriorated} transparency as a three-dimensional aberrating and scattering structure.
%\begin{figure*}[htbp]
%\includegraphics[width=0.6\textwidth]{figure3}
%\caption{\textbf{Imaging through a thin aberrative layer}. (a) Schematic of the experiment. A resolution target (USAF 1951) is positioned at a distance $d = 2$  mm underneath a rough platic film (see inset). (b-d) Distortion phase masks associated with the first three singular values.(e-g) Singular spaces in the image plane associated to the first three singular values. Comparison of the optical coherence tomography image of the focal plane without correction (h), with a global correction using the first distortion phase mask  (i)and with the local correction (j). Scalebar 50 microns.}
%\label{fig4}
%\end{figure*}

\subsection*{Imaging through a hazy cornea}

\begin{figure*}[htbp]
\center
\includegraphics[width=\textwidth]{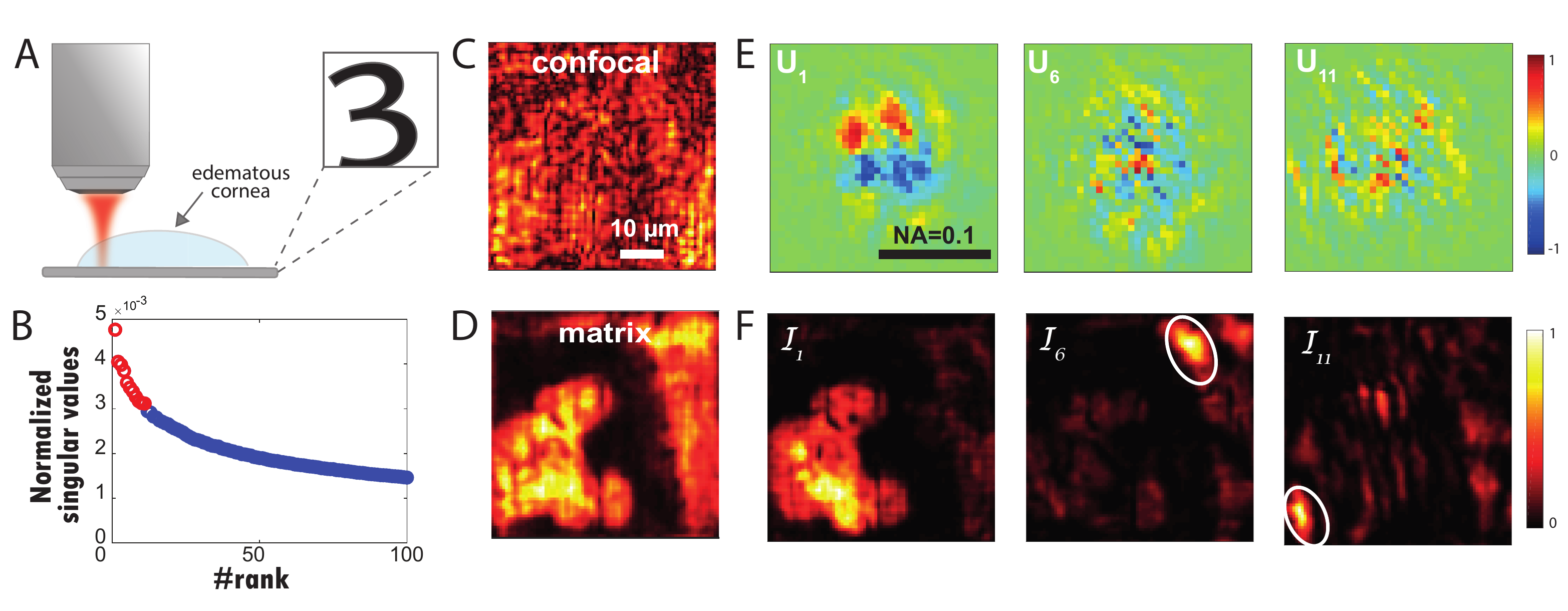}
\caption{\textbf{Imaging \cora{through corneal tissue with deteriorated transparency}}. (\textbf{A}) Schematic of the experiment. A resolution target (USAF 1951) is positioned below \cora{an edematous} non-human primate cornea (see inset). (\textbf{B}) Plot of the normalized singular values $\tilde{\sigma}_i$ of $\mathbf{D}$. The red circles correspond to the eleven first singular values (signal subspace), while the noisy singular values are displayed in blue. (\textbf{C}) Original confocal image deduced from the focused reflection matrix $\mathbf{R_0}$ (Eq.~\ref{I}). (\textbf{D}) Final matrix image constructed from the eleven first eigenstates of $\mathbf{D}$ (Eq.~\ref{I2}). (\textbf{E}) Real parts of $\mathbf{U_1}$, $\mathbf{U_6}$ and $\mathbf{U_{11}}$. (\textbf{F}) Corresponding confocal images deduced from the focused reflection matrices $\mathbf{R_p}$ (Eq.~\ref{I}).}
\label{fig5}
\end{figure*}
The experimental configuration is displayed in Fig.~\ref{fig5}A. The number ``3'' of the group 5 in the resolution target is imaged through a 700-$\mu$m-thick edematous non-human primate cornea. 
The reflection matrix $\mathbf{R}$ is measured over a FOI $52 \times 52$ $\mu$m$^2$ by means of $N_\textrm{in}$=625 input wave-fronts, a spatial sampling $\delta r_\textrm{in}=2.1$ $\mu$m and an input pupil aperture $\mathcal{D}_\textrm{in}\times \mathcal{D}_\textrm{in}=1\times1$ mm$^2$. At the output, the wave-field is recorded over an output pupil size $\mathcal{D}_\textrm{out} \times \mathcal{D}_\textrm{out} =2 \times 2$ mm$^2$ with $N_\textrm{out}=1296$ pixels and a spatial sampling length $\delta u_\textrm{out}$=56 $\mu$m. Fig.~\ref{fig5}C displays the confocal image $\mathcal{I}_0$ deduced from $\mathbf{R_0}$ with $l_p=1$ $\mu$m (Eq.~\ref{I}). Multiple scattering and aberrations induced by the cornea induce a random speckle reflected wave-field that prevents from imaging the resolution target. On the contrary, as we will see, the $\mathbf{D}$-matrix analysis allows us to nicely recover the pattern ``3'' of the resolution target (see Fig.~\ref{fig5}D).  

Fig.~\ref{fig5}C displays the spectrum of the singular values $\tilde{\sigma}_i$. The first singular value emerges from the rest of the spectrum but it is difficult to know until which rank the eigenstates can be considered as belonging to the signal subspace. As previously, the Shannon entropy  of the singular values yields an unambiguous answer: $\mathcal{H}(\sigma_i)=10.7$. The 11$^{th}$ first singular states should thus be considered.  Fig.~\ref{fig5}E displays the 1$^{st}$, 6$^{th}$ and 11$^{th}$ singular vectors $\mathbf{U_i}$ in the pupil plane. The complexity of the wave-front distortion, \textit{i.e} their spatial frequency content, increases with the rank of the corresponding singular values. The corresponding IMs $\mathcal{I}_p$(Eq.~\ref{I}) are displayed in Fig.~\ref{fig5}F. While the first singular vector $\mathbf{U_1}$ allows a wide-field correction of low-order aberration, the higher rank singular vectors are associated with high-order aberrations that are effective over IMs of smaller dimension. In Fig.~\ref{fig5}D, the whole spatial frequency spectrum of wave-front distortions is finally compensated by smartly combining the confocal images $\mathcal{I}_p$ associated with each singular state from $\mathbf{D}$'s signal subspace (Eq.\ref{I2}). The comparison of the initial (Fig.~\ref{fig5}C) and final (Fig.~\ref{fig5}D) images is spectacular with a Strehl ratio gain $\mathcal{S}'_1/S'_0=230$. The comparison of $\mathcal{I}$ (Fig.~\ref{fig5}D) and $\mathcal{I}_1$ (see the first inset of Fig.~\ref{fig5}F) illustrates the benefit of a matrix approach of aberration correction compared to conventional AO, since the latter one would yield $\mathcal{I}_1$ in theory. 

This decomposition of complex aberration phase laws over a set of IMs opens important perspectives for aberrometry. It actually goes well beyond state-of-the-art techniques that basically consist in a simple projection over a set of Zernike polynomials. Moreover, an estimator of the single-to-multiple scattering (SMR) ratio can be built on the relative energy between the signal and noise sub-spaces of $\mathbf{D}$:
\begin{equation}
\mbox{SMR}=\frac{{\sum_{i=1}^{\mathcal{H}(\sigma_i)} \sigma_i^2}}{{\sum_{i=\mathcal{H}(\sigma_i)+1}^{N_\textrm{in}}} \sigma_i^2}.
\label{eq_SMR}
\end{equation}
The SMR can actually be a quantitative bio-marker of the \cora{corneal opacification or a quantitative measure of corneal transparency~\cite{Bocheux:19}}. \cora{Based on a fit with a recent analytical study of the SMR~\cite{badon2017multiple}, the cornea thickness $L$ can be estimated in terms of scattering mean free path $\ell_s$: $L \sim 9 \ell_s$ (see Fig.~S3). As the corneal thickness is known ($L=700$ $\mu$m), the scattering mean free path can be deduced:  $\ell_s \sim 80$ $\mu$m. Interestingly, this value is in excellent agreement with recent \textit{ex-vivo} measurements of $\ell_s $ in pathological corneas with compromised transparency~\cite{Bocheux:19}. The value of 9$\ell_s$ highlights the difficult experimental conditions under which the imaging of the resolution target has been successfully achieved.} 

In conclusion, this last experiment shows the potential of a matrix approach for eye aberrometry and \cora{turbidimetry, such as for improved quality control of donor tissue assessment prior to corneal transplantation~\cite{Bocheux:19}.} Of course, this method is by no means limited to \cora{ophtalmic applications}. It can be applied to the characterization of any kind of biological tissues provided that we have access to the associated reflection matrix.

\section*{DISCUSSION}
In this article, we present a novel and non-invasive method for aberration compensation and diffraction-limited imaging at large optical depths. \cora{This approach relies on a new operator, the so-called distortion matrix, that connects a set of input focusing points with the distorted component of the reflected the wave-field  in the pupil plane.} \coraa{This operator connecting position and spatial frequency has some analogy with the Wigner distribution function~\cite{bastiaans1979wigner}.} \cora{However, the Wigner transform applies to a single variable of a function, \textit{i.e} to a single vector in a discrete formalism. Here, our position-momentum analysis is performed between the input and output of a reflection matrix.}

\cora{The concept of distortion matrix is to measure the back-scattered waves in a de-scanned frame while scanning the sample with focused illuminations. This approach has some similarity with a previous AO approach~\cite{rueckel2006adaptive} in its hardware configuration. The main difference is that, in this study, wave-fronts are averaged by the Shack-Hartmann type of analysis and \cora{this AO approach} thus relies on an isoplanatic condition.}
\coraa{Here lies one of the strengths of our approach. While conventional methods estimate the aberrated wavefront for a single location or averaged over the whole FOI, we propose to study the spatial and angular correlations of the distortion operator through an SVD. In this manner, we demonstrate the efficient compensation of both low- and high-order aberrations over multiple IPs. Moreover, our approach relies on the Shannon entropy that provides an objective criterion to determine the number of IPs supported by the FOI. This is in contrast with recent works \cora{based on a far-field reflection matrix} in which the FOI was arbitrarily divided into sub-areas where different corrections were applied \cite{kim2019label,yoon2019laser}.}

Besides aberration correction, our approach leverages the correlations of the output wave-field to get rid of the multiple scattering background. The latter contribution is actually spatially incoherent. It thus mainly lies along the noise subspace of the $\mathbf{D}$-matrix. Thanks to these features, we were able to image through \cora{almost} 10 scattering mean free paths of biological tissues, which is beyond the imaging depth of conventional OCT systems \coraa{for such specimens (see Fig.~S3)}. Compared to the previously developed smart-OCT method that was able to detect few bright scatterers at large penetration depth (12$\ell_s$)~\cite{badon2016smart}, the $\mathbf{D}$-matrix approach yields a direct image of the sample reflectivity at a diffraction-limited resolution. \coraa{Additionally, our approach enables to quantitatively estimate the amount of multiply-scattered light. Combined with a conventional image, this parameter is of importance for characterization purposes.}

The distortion operator thus opens a new route towards real-time deep optical imaging of biological tissues. In that respect, the experimental set up and procedure used in this paper are clearly perfectible. \coraa{While post-process operations take less than one minute on a regular laptop, the main limitation in the current experimental configuration is the acquisition time. In particular, the scanning illumination scheme was not optimized because of the SLM speed. While the use of a galvanometric mirror or a high-speed deformable mirror would reduce drastically the acquisition time at the cost of a more complex setup, we counteracted this issue with a sparse illumination. However, this, in return, limited the available number of angular degrees of freedom at the input, which prevents us from an aberration correction }\cora{of the incident wave-field. By optimizing the experimental apparatus and acquisition scheme, large reflection matrices can be measured in a few seconds. For instance, Yoon \textit{et al.} recently demonstrated the acquisition of a 10000 modes matrix in 15 seconds \cora{with the same degree of control for the incident and reflected waves}~\cite{yoon2019laser}}. \cora{In that case, a simultaneous correction of aberrations at the input and output is absolutely possible under the distortion matrix approach by alternatively projecting the incident and reflected wave-fields in the focal and pupil planes. In view of 3D imaging, our approach can also be coupled to computational AO techniques~\cite{adie2012computational} in order to tackle depth-dependent aberrations and restore a diffraction-limited resolution in all directions. An alternative is to switch from a scanning to a full-field illumination scheme. A measurement of the coherent reflection matrix $\mathbf{R}$ can be performed under a spatially incoherent illumination \cite{PhysRevLett.114.023901,badon2016spatio}. This full-field configuration would allow to record the reflection matrix over millimetric volumes in a moderate acquisition time.} 

\coraa{Finally, we used a negative resolution target as the sample to be imaged in this work. The reason is that this highly contrasted object was the ideal specimen to clearly highlight the issue of multiple isoplanatic areas. Beyond the proof-of-concept experiments presented in this article, a direct imaging of biological specimens over large penetration depth will be the next step. \cora{Interestingly, the assumption on which our method is based (Eq.~\ref{strong_ab}) can easily be met in biological tissues since a strong aberration regime takes place beyond a few scattering mean free paths. Note also that, even when this condition is not fulfilled and far-field correlations dominate, the distortion matrix approach can still work but the FOI has to be beforehand sub-divided into individual IPs~\cite{kim2019label,Lambert2020}.} The ability of identifying multiple IPs will also be particularly promising to map the specimen-induced aberration \cora{and the single-to-multiple scattering ratio}. \cora{Aside from aberrometry and/or turbidimetry, future in-vivo implementations of our approach have implications beyond that of ocular media characterization, most notably for imaging through non-transparent ocular media (e.g., retinal imaging through a turbid cornea or through cataract opacities)~\cite{Badon2019}.}}

In summary, we have introduced, in this work, a new operator, the so-called distortion matrix $\mathbf{D}$, which reveals the hidden correlations of the reflected wave-field. This matrix results from the mismatch between the phase of the recorded reflection matrix and those of a reference matrix that would be obtained in an ideal configuration. As shown in this paper, $\mathbf{D}$ gives access to the non-invasive transmission matrix between each sensor and each voxel of the \cora{FOI}. Then, by solving the corresponding inverse problem, an image of a scattering sample can be obtained as if the medium ahead was made transparent. The $\mathbf{D}$-matrix concept is very general. It can be extended to any kind of waves and experimental configurations for which a measurement of the amplitude and phase of the reflected wave-field is possible under multiple illuminations~\cite{Lauer,shahjahan,Blondel2018,Zhang:18}. \coraa{A recent work actually demonstrates the benefits of this concept for ultrasound imaging in a random scattering regime~\cite{Lambert2020}}. This $\mathbf{D}$-matrix concept thus opens a new route towards a global and non-invasive matrix approach of deep imaging in biological tissues.

\section*{MATERIALS AND METHODS}
\subsection*{Experimental set up}
The experimental configuration is identical to the one described in \cite{badon2016smart} except for the MO that had been replaced by a water immersion one. The following components were used in the experimental setup (see Fig.~S1): a femtosecond laser (Femtosecond Fusion 20-400, \coraa{central wavelength: 810 nm, bandwidth: 40 nm}), an SLM (PLUTONIR-2, HOLOEYE), an objective lens (40$\times$; NA, 0.8; Nikon), and a CCD camera (Dalsa Pantera 1M60) with a dynamic range of 60 dB. The incident light power in the back pupil plane of the MO was 10 mW in the experiment. Thus, the radiant flux was $10^6$ W/cm$^2$ at the focal spot in free space. For each input wave-front, the complex-reflected wave field was extracted from four intensity measurements using phase shifting interferometry. The acquisition time of the reflection matrix was approximately 2 minutes.

\subsection*{Image acquisition and data analysis}
Both data acquisition and analysis were performed using Matlab custom-written codes. These codes are available from the authors upon request.

\section*{SUPPLEMENTARY MATERIALS}

\cora{Section~S1. Correlations of the reflected and distorted wave-fields in the pupil plane.}

\noindent \cora{Section~S2. Correlations of the reflected and distorted wave-fields in the focal plane.}

\noindent \cora{Section~S3. Singular value decomposition of the distortion matrix.}

%\noindent Section~\ref{suppMS}. Theoretical estimation of the single-to-multiple scattering ratio

\noindent Fig. S1. Measuring the time-gated reflection matrix.

\noindent \cora{Fig. S2. Conjugating the pupil, focal and imaging planes.}

\noindent \cora{Fig. S3. Predicting the single-to-multiple scattering ratio in biological tissues.}

\noindent Fig. S4. Building the reflection matrix $\mathbf{R}$.

\noindent Fig. S5. Modeling light propagation from the virtual source plane to the output pupil plane.

\noindent \cora{Tab. S1. Glossary of the variables used in the study.}

\noindent \cora{Tab. S2. Glossary of the matrices used in the study.}

\noindent \cora{References~\cite{Robert_thesis,Priestley1988,Goodman2000,Prada2003,prada1994eigenmodes}.}

\section*{REFERENCES AND NOTES}

%\bibliography{nphoton}
\bibliographystyle{ScienceAdvances}

\subsection*{Acknowledgements} 
\cora{The authors wish to thank Laura Cobus, William Lambert, Paul Balondrade and Serge Meimon for fruitful discussions.} 
\noindent \textbf{Funding:} 
The authors are grateful for the funding provided by Labex WIFI (Laboratory of Excellence within the French Program Investments for the Future) (ANR-10-LABX-24 and ANR-10-IDEX-0001-02 PSL*). A.B. acknowledges financial support from the French ``Direction G\'{e}n\'{e}rale de l'Armement'' (DGA). This project has received funding from the European Research Council (ERC) under the European Union's Horizon 2020 research and innovation programme (grant agreements nos. 610110 and 819261, HELMHOLTZ*
and REMINISCENCE projects, respectively). K.I. acknowledges financial support from the European Union's Horizon 2020 research and innovation programme under the Marie Sk\l{}odowska-Curie grant agreement No. 709104.\\
\noindent \textbf{Author Contributions} A.A. initiated and supervised the project. A.B. built the experimental setup and performed the experiments. \cora{K.I., A.C.B and M.F. initiated the ophthalmic application. K.I. provided corneal samples and guidance on the ophthalmic experiment.} A.B., V.B., and A.A. analyzed the experiments. V.B. and A.A. performed the theoretical study. A.B. and A.A. prepared the manuscript. A.B., V.B., K.I., A.C.B., M.F., and A.A. discussed the results and contributed to finalizing the manuscript.\\
\noindent \textbf{Competing Interests:} \cora{A.A., M.F., A.C.B, A.B. and V.B. are inventors on a patent related to this work held by CNRS (no. WO2020016249, published January 2020). All authors declare that they have no other competing interests.}\\
\noindent \textbf{Data and materials availability:} All data needed to evaluate the conclusions in the paper are present in the paper and/or the Supplementary Materials. Additional data related to this paper may be requested from the authors.

\clearpage 

\clearpage

\renewcommand{\thetable}{S\arabic{table}}
\renewcommand{\thefigure}{S\arabic{figure}}
\renewcommand{\theequation}{S\arabic{equation}}
\renewcommand{\thesection}{S\arabic{section}}
\renewcommand{\thesubsection}{S\arabic{section}.\arabic{subsection}}

\setcounter{equation}{0}
\setcounter{figure}{0}
\setcounter{page}{1}

\newpage

\section*{Supplementary Materials}

\section{\label{angular_corr}Correlations of the reflected and distorted wave-fields in the pupil plane}

In this section, we derive the pupil correlations of the $\mathbf{R}$- and $\mathbf{D}$- matrices. Our aim is to provide a theoretical proof of the experimental observation made in Fig.\ref{fig1}C2 and \ref{fig1}D2. The distorted wave-fields exhibit correlations over a longer range than the reflected wave-fields in the pupil plane. For sake of analytical tractability but without loss of generality, we will assume in this section: (\textit{i}) a set of fully incoherent input focal spots (\textit{i.e} a strong aberration regime);  (\textit{ii}) a field-of-illumination (FOI) contained in a single IP. The main result is the following: While the pupil correlation length $r_P$ of $\mathbf{R}$ scales as the inverse of the FOI size ($r_P\sim \lambda f/\Omega$), the correlation length $d_P$ of $\mathbf{D}$ is inversely proportional to the width $\delta_\textrm{in}$ of the input PSF ($r_P\sim \lambda f/\delta_\textrm{in}$). The proofs of these two assertions are provided below.

\subsection{Reflection matrix}

To investigate the angular correlations of the reflected wave-field, the correlation matrix \alex{$ \mathbf{B_R}=N_\textrm{in}^{-1}\mathbf{R} \mathbf{R}^{\dag}$} should be considered. Using Eq.~\ref{int_abe2}, its coefficients can be expressed as follows:\alex{
\begin{equation}
\label{B1}
 B_R (\mathbf{u}_\mathrm{out},\mathbf{u'}_\mathrm{out})  = N_\textrm{in}^{-1} \int_{\Omega} d\mathbf{r}  \int_{\Omega}  d\mathbf{r'} T(\uout,\mathbf{r}) T^*(\upout,\mathbf{r}^{\prime})  \gamma(\mathbf{r}) \gamma^*(\mathbf{r'}) \sum_{\rin}  H_\textrm{in}(\mathbf{r},\rin)H^*_\textrm{in}(\mathbf{r'},\rin).
\end{equation}}
In a strong aberration regime, the input focal spots can be considered as fully incoherent, 
\begin{equation}
\label{random}
 \left \langle  H_\mathrm{in}(\mathbf{r},\rin) H_\mathrm{in}^*(\mathbf{r'},\rin) \right \rangle  = \left \langle \left | H_\mathrm{in}(\mathbf{r},\rin) \right |^2 \right \rangle
\delta (\mathbf{r}-\mathbf{r'}).
\end{equation}
where  $\delta$ is the Dirac distribution and the symbol $\langle \cdots \rangle$ denotes an ensemble average. In a strong aberration regime, $\mathbf{B_R}$ can be decomposed as the sum of a covariance matrix $\left \langle \mathbf{B_R}\right \rangle $ and a perturbation term $\delta  \mathbf{B_R} $:
\begin{equation}
\label{B}
   \mathbf{B_R}= \left \langle \mathbf{B_R}\right \rangle+\delta  \mathbf{B_R} ,
\end{equation}
  The correlation matrix $\mathbf{B_R}$ (Eq.~\ref{B}) should converge towards the covariance matrix $\left \langle \mathbf{B}\right \rangle$ for a  sufficiently large number $M_R \sim (\Omega/\delta^\textrm{0}_\textrm{in})^2$ of independent speckle grains in the focal plane (Eq.\ref{MR}). More precisely, the intensity of the perturbation term in Eq.\ref{B}, $|\delta \mathbf{B_R}|^2$, scales as the inverse of $M_R$ ~\cite{Robert_thesis,Priestley1988,Goodman2000}. 

Assuming the convergence of $\mathbf{B_R}$ towards $\langle \mathbf{B_R} \rangle $ ($M_R>>1$), the correlation coefficients $B_R(\uout,\upout)$ (Eq.~\ref{B1})  can be expressed as follows: \alex{
\begin{eqnarray}
 B_R (\uout,\upout) = N_\textrm{in}^{-1} \int d\mathbf{r} T(\uout,\mathbf{r}) |\gamma(\mathbf{r})|^2 T^*(\upout,\mathbf{r}) \times  \sum_{\rin} \left \langle \left | H_\mathrm{in}(\mathbf{r},\rin) \right |^2 \right \rangle,
 \label{B2}
\end{eqnarray}}
To go further, an isoplanatic configuration should be considered. On the one hand, this means that the input PSF is invariant by translation:
\begin{equation}
\label{iso1}
    H_\textrm{in}(\mathbf{r},\rin)=H_\textrm{in}(\mathbf{r}-\rin)
  \end{equation}
On the other hand, the output transmission matrix coefficients $T(\uout,\mathbf{r})$ can be decomposed as the product of the transmittance $\hat{H}_\textrm{out}(\uout)$ of the aberrating layer and the free-space transmission matrix coefficients $T_0(\uout,\mathbf{r})$ (Eq.\ref{G0}):
\begin{equation}
\label{iso2}
    T(\uout,\mathbf{r})=\hat{H}_\textrm{out}(\uout) T_0(\uout,\mathbf{r}).
    \end{equation}
Injecting these last equations and Eq.~\ref{G0} into Eq.~\ref{B2} leads to the following expression for :
\begin{equation}
\label{B8}
    B_R (\uout,\upout)= I_\textrm{in} \hat{H}_\textrm{out}(\uout) \hat{H}_\textrm{out}^*(\upout)  \hat{\gamma} (\upout-\uout)   
\end{equation}
where \alex{$$I_\textrm{in}=N_\textrm{in}^{-1}\sum_{\rin} \left \langle \left | H_\mathrm{in}(\mathbf{r}-\rin) \right |^2 \right \rangle$$} is the mean input PSF intensity and $$\hat{\gamma}(\mathbf{u})=\int d\mathbf{r} |\gamma(\mathbf{r})|^2 \exp (-j 2\pi \mathbf{u}.\mathbf{r}/\lambda f)$$ is the 2D Fourier transform of the scattering distribution $|\gamma(\mathbf{r})|^2$ in the focal plane. This quantity, which dictates the correlations displayed by $\mathbf{R}$ in the pupil plane, can be seen as an incoherent structure factor of the object placed in the FOI. The corresponding coherence length $r_p$ scales as
\begin{equation}
\label{thetaR}
r_P \sim \lambda f / \Omega ,
\end{equation}
The number $N_R$ of independent speckle grains in the reflected wave-field is given by the squared ratio between the output pupil size $\mathcal{D}_\textrm{out}=\lambda f / \delta_\textrm{out}^\textrm{0}$ and the pupil coherence length $r_P$: 
\begin{equation}
\label{NR}
    N_R \sim ( {\Omega}/{\delta_\textrm{out}^\textrm{0}})^2
\end{equation}
$N_R$ scales as the number of output resolution cells mapping the object.

These theoretical predictions account for the incoherence of the reflected wave-field shown in Fig.~\ref{fig1}C3. This figure plots the auto-correlation function $\mathcal{B}_R(\Delta \mathbf{u}) $ of the reflected wave-field in the pupil plane. It is computed by averaging the correlation matrix coefficients $B_R(\uout,\upout)$ over couples $(\uout,\upout)$ sharing the same relative position $\Delta \mathbf{u}=\uout -\upout$. 

\subsection{Distortion matrix}

As highlighted by Fig.~\ref{fig1}C and demonstrated above, the reflection matrix displays a random feature at the output in the strong aberration regime. Now we will show how the realignment of the reflected wave-fronts in the pupil plane can reveal the angular correlations of the distorted component.

The distortion matrix $\mathbf{D}$ is defined as the Hadamard product between the reflection matrix $\mathbf{R}$ and the reference transmission matrix $\mathbf{T}_0^{*}$ (Eqs.~\ref{D}-\ref{D2}). In the isoplanatic limit (Eqs.~\ref{iso1}-\ref{iso2}) and using Eq.\ref{int_abe2}, the $\mathbf{D}$-matrix coefficients can be expressed as follows
\begin{equation}
\label{Dkr}
D(\uout,\rin)= \hat{H}_\textrm{out}(\uout) \int d\mathbf{r}  T_0(\uout,\mathbf{r}-\rin) \gamma(\mathbf{r}) H_\textrm{in}(\mathbf{r}-\rin) .
\end{equation}
To investigate the angular correlations between distorted wave-fields, the spatial correlation matrix \alex{$\mathbf{B_D} = N_\textrm{in}^{-1}\mathbf{D} \mathbf{D}^{\dag}$} is investigated. Its coefficients can be expressed as follows:\alex{
\begin{eqnarray}
\label{C1}
B_D (\mathbf{u}_\mathrm{out},\mathbf{u'}_\mathrm{out})   & = & N_\textrm{in}^{-1} \hat{H}(\uout) \hat{H}^*(\upout)    \\
 & \times &    \int d\mathbf{r_1}  \int  d\mathbf{r_2}  \gamma(\mathbf{r_1})  \gamma^*(\mathbf{r_2}) \nonumber \\
&\times & \sum_{\rin} H_\mathrm{in}(\mathbf{r_1}-\rin) T_0(\uout,\mathbf{r_1}-\rin) H_\mathrm{in}^*(\mathbf{r_2}-\rin) T_0^*(\upout,\mathbf{r_2}-\rin)\nonumber
\end{eqnarray}}
As $\mathbf{B_R}$ (Eq.~\ref{B}), $\mathbf{B_D}$ can  be decomposed as the sum of a covariance matrix $\left \langle \mathbf{B_D}\right \rangle $ and a perturbation term $\delta  \mathbf{B_D} $ whose intensity decreases with the number $M_D \sim (\Omega/\ell_F)^2$ of independent speckle grains for the distorted wave-field from the focal plane (Eq.~\ref{MD}). For $M_D>>1$, $\mathbf{B_D}$ converges towards $\left \langle \mathbf{B_D}\right \rangle$, such that:\alex{
\begin{eqnarray}
\label{C1b}
B_D (\mathbf{u}_\mathrm{out},\mathbf{u'}_\mathrm{out})   &= & N_\textrm{in}^{-1}\hat{H}(\uout) \hat{H}^*(\upout) \\
 & \times & \int d\mathbf{r_1}  \int  d\mathbf{r_2}  \gamma(\mathbf{r_1})  \gamma^*(\mathbf{r_2}) \nonumber \\
 & \times &  \sum_{\rin} \langle H_\mathrm{in}(\mathbf{r_1}-\rin) H_\mathrm{in}^*(\mathbf{r_2}-\rin) \rangle T_0(\uout,\mathbf{r_1}-\rin)  T_0^*(\upout,\mathbf{r_2}-\rin)\nonumber
\end{eqnarray}} 
Assuming a strong aberration regime (Eq.~\ref{random}), the expression of the correlation matrix coefficients $B_D(\uout,\upout)$can be simplified as follows 
\begin{equation}
\label{C2}
B_D (\mathbf{u}_\mathrm{out},\mathbf{u'}_\mathrm{out})  = I_0 \hat{H}(\uout) \hat{H}^*(\upout) \int d\mathbf{r_1} | \gamma(\mathbf{r_1}) |^2 \sum _{\mathbf{r}^{\prime}_\textrm{in} }   T_0(\uout,\mathbf{r}^{\prime}_\textrm{in}) T_0^*(\upout,\mathbf{r}^{\prime}_\textrm{in})   \gamma_D(\mathbf{r}^{\prime}_\textrm{in})
\end{equation}
with $\mathbf{r}^\prime_\textrm{in}=\mathbf{r_1}-\rin$ and \alex{
\begin{equation}
\label{gammaD}
\gamma_D(\mathbf{r}^{\prime}_\textrm{in})=\left \langle | H_\mathrm{in}(\mathbf{r}^{\prime}_\textrm{in})|^2 \right \rangle , 
\end{equation}}
the intensity distribution of the virtual source synthesized in the focal plane at the input. Using Eqs.~\ref{G0} and \ref{iso2}, Eq.~\ref{C2} can be rewritten as 
\begin{equation}
\label{C3}
B_D (\mathbf{u}_\mathrm{out},\mathbf{u'}_\mathrm{out})  \propto  \hat{H}(\uout) \hat{H}^*(\upout)  \hat{\gamma}_D (\upout-\uout)
\end{equation}
where $\hat{\gamma}_D(\mathbf{u})=\sum_{\mathbf{r}} \gamma_D(\mathbf{r}) \exp (-j 2\pi \mathbf{u}.\mathbf{r}/\lambda f)$ is a discrete 2D Fourier transform of the scattering distribution $\gamma_D(\mathbf{r})$ in the focal plane. The correlation length $d_p$ of the distorted wave-field in the pupil plane is thus inversely proportional to the spatial extension $\delta_\textrm{in}$ of the input PSF intensity $| H_\mathrm{in}|^2$, such that
\begin{equation}
\label{thetaD}
d_P \sim \lambda f / \delta_\textrm{in}.
\end{equation}
The number of independent speckle grains in the distorted wave-field is the squared ratio between 
the output pupil size $\mathcal{D}_\textrm{out}=\lambda f / \delta_\textrm{out}^\textrm{0}$ and the pupil coherence length $d_P$: 
\begin{equation}
\label{ND}
    N_D \sim ({\delta_\textrm{in}}/{\delta_\textrm{out}^\textrm{0}})^2
\end{equation}
$N_D$ scales as the number of output resolution cells mapping the input PSF.

As $\delta_\textrm{in}$ is smaller than the FOI dimension $\Omega$, $d_P$/$N_D$ are larger/smaller than $r_P$/$N_R$ (Eqs.~\ref{thetaR}-\ref{NR}), respectively.  This highlights the enhancement of the far-field correlations in $\mathbf{D}$ shown in Fig.~\ref{fig1}D3. This figure plots the auto-correlation function $\mathcal{B}_D(\Delta \mathbf{u}) $ of the distorted wave-field in the pupil plane. $\mathcal{B}_D(\Delta \mathbf{u}) $ is computed by averaging the correlation matrix coefficients $B_D(\uout,\upout)$ over couples $(\uout,\upout)$ of common relative position $\Delta \mathbf{u}=\uout -\upout$.

\section{\label{spatial_corr}Spatial correlations of the reflected and distorted wave-fields}

In this section, we derive the input correlations of the matrices $\mathbf{R}$ and $\mathbf{D}$. Our aim is to provide a theoretical proof of the experimental observation made in Fig.\ref{fig1}C2 and \ref{fig1}D2. As seen previously in the pupil plane, the distorted wave-fields reveal spatial correlations in the focal plane that were originally hidden in the recorded wave-fields. Unlike the previous section, we derive a general expression for the input correlation matrices beyond the isoplanatic limit. The main result is the following: While the correlation length $r_F$ of the reflected wave-field in the focal plane is restricted to the input diffraction limit resolution $\delta_\textrm{in}^\textrm{0}$, the correlation length $d_F$ of $\mathbf{D}$ in the focal plane corresponds to the isoplanatic length $\ell_c$.  

\subsection{Reflection matrix}

To investigate the spatial correlations of the reflected wave-field, the correlation matrix $ \mathbf{C_R}={N^{-1}_\textrm{out}}\mathbf{R}^{\dag}\mathbf{R}$ should this time be considered. Unlike in the previous section, the isoplanatic assumption is here not made.  
Using Eq.~\ref{int_abe2}, the coefficients of $ \mathbf{C_R}$ can be expressed as follows:
\begin{equation}
\label{CC1}
 C_R (\mathbf{r}_\mathrm{in},\mathbf{r'}_\mathrm{in})  =  {N^{-1}_\textrm{out}}\int d\mathbf{r}  \int  d\mathbf{r'}  \gamma(\mathbf{r}) \gamma^*(\mathbf{r'})   H_\textrm{in}(\mathbf{r},\rin)H^*_\textrm{in}(\mathbf{r'},\rpin)\sum_{\uout}  T(\uout,\mathbf{r})   T^*(\uout,\mathbf{r'}) 
\end{equation}
As correlation matrices in the pupil plane, $\mathbf{C_R}$ converges towards the covariance matrix $\langle \mathbf{C_R} \rangle$ for a large number $N_R\sim(\Omega/\delta^{0}_\textrm{out})^2$ of independent speckle grains for the reflected wave-field in the pupil plane (Eq.~\ref{MR}). 
For $N_R>>1$, the coefficients of $\mathbf{C_R}$ are given by:
\begin{equation}
\label{CC2b}
 C_R (\mathbf{r}_\mathrm{in},\mathbf{r'}_\mathrm{in})  =  {N^{-1}_\textrm{out}}\int d\mathbf{r}  \int  d\mathbf{r'}  \gamma(\mathbf{r}) \gamma^*(\mathbf{r'})   H_\textrm{in}(\mathbf{r},\rin)H^*_\textrm{in}(\mathbf{r'},\rpin)\sum_{\uout} \left \langle T(\uout,\mathbf{r})   T^*(\uout,\mathbf{r'}) \right \rangle
\end{equation}
The mean correlation term $\langle T(\uout,\mathbf{r})   T^*(\uout,\mathbf{r'}) \rangle$ can be developed by writing the transmission matrix as a Hadamard product between the free-space transmission matrix $\mathbf{T_0}$ and an aberration matrix $\mathbf{H_{\textrm{out}}}$, such that $$T(\uout,\mathbf{r})= \hat{H}_{\textrm{out}}(\uout,\mathbf{r})T_0(\uout,\mathbf{r}).$$ It comes\alex{
\begin{eqnarray}
  \left  \langle T(\uout,\mathbf{r})   T^*(\uout,\mathbf{r'}) \right \rangle & = & \left  \langle \hat{H}_\mathrm{out}(\uout, \mathbf{r}) \hat{H}_\mathrm{out}^*(\uout, \mathbf{r'}) \right \rangle T_0(\uout,\mathbf{r})T_0^*(\uout,\mathbf{r'})
  \\
   & = & F(\mathbf{r},\mathbf{r'}) \left  \langle \left | \hat{H}_\mathrm{out}(\uout, \mathbf{r}) \right |^2 \right \rangle T_0(\uout,\mathbf{r})T_0^*(\uout,\mathbf{r'}).
   \label{corrT}
\end{eqnarray}}
The correlation function, 
\begin{equation}
\label{FF}
F(\mathbf{r},\mathbf{r'})=\left  \langle \hat{H}_\mathrm{out}(\uout, \mathbf{r}) \hat{H}_\mathrm{out}^*(\uout, \mathbf{r'})   \right \rangle / \left  \langle \left | \hat{H}_\mathrm{out}(\uout, \mathbf{r}) \right |^2 \right \rangle,
\end{equation}
describes the spatial correlation of the aberration matrix $\mathbf{\hat{H}}_\mathrm{out}$ in the focal plane. Its support is directly related to $\ell_c$, the IP size. For sake of simplicity but without lack of generality, we assume that the aberrating layer does not attenuate the wave-field: 
\begin{equation}
\label{att}
\left  \langle \left | \hat{H}_\mathrm{out}(\uout, \mathbf{r}) \right |^2 \right \rangle=1.
\end{equation}
Using Eq.~\ref{corrT}, the sum over $\uout$ into Eq.~\ref{CC2b} can then be rewritten as:
\begin{equation}
  {N^{-1}_\textrm{out}}\sum_{\uout} \left \langle T(\uout,\mathbf{r})   T^*(\uout,\mathbf{r'}) \right \rangle  =  F(\mathbf{r},\mathbf{r'})  \sum_{\uout} T_0(\uout,\mathbf{r})T_0^*(\uout,\mathbf{r'})
    \end{equation}
Injecting the expression of the coefficients ${T_0}(\uout,\rin)$ (Eq.~\ref{G0}), it finally comes
    \begin{eqnarray}
{N^{-1}_\textrm{out}}\sum_{\uout} \left \langle T(\uout,\mathbf{r})   T^*(\uout,\mathbf{r'}) \right \rangle & = & F(\mathbf{r},\mathbf{r'})  \sum_{\uout} \exp \left ( i \frac{2\pi}{\lambda f} \uout . (\mathbf{r}-\mathbf{r'})\right ) \nonumber \\
& = &   \delta  (\mathbf{r}-\mathbf{r'}) 
\end{eqnarray}
The physical meaning of this last equation is that two virtual sources located at points $\mathbf{r}$ and $\mathbf{r'}$ in the focal plane give rise to uncorrelated wave-fields in the pupil plane. Injecting this last relation into Eq.~\ref{CC2b} leads to the following expression for $C_R(\rin,\rpin)$
 \begin{equation}
\label{CC2}
  C_R (\mathbf{r}_\mathrm{in},\mathbf{r'}_\mathrm{in})  =  \int d\mathbf{r} 
 | \gamma(\mathbf{r})|^2 H_{\textrm{in}}(\mathbf{r},\rin)H_\textrm{in}^*(\mathbf{r},\rpin)
\end{equation}
%To go further, the characteristic length scale $l_{\rho}$ of the reflectivity's square norm is assumed to be larger than the length scale of the input focal spot's fluctuation. Under this condition,  Eq.~\ref{C2} can be simplified into
%\begin{equation}
%\label{C3}
 %C (\mathbf{r}_\mathrm{in},\mathbf{r'}_\mathrm{in})  =  \int d\mathbf{r} 
 %| \rho(\mathbf{r})|^2 H_{\textrm{in}}(\mathbf{r}-\rin)H_\textrm{in}^*(\mathbf{r}-\rpin)
%\end{equation} a
%To assess the mean spatial correlation of the reflected wave-field, a statistical average of the last equation should be considered. 
%The coefficients of the corresponding covariance matrix $ \langle \mathbf{C_R} \rangle$ can be expressed as
 %\begin{equation}
%\label{CC3}
 %\langle C_R (\mathbf{r}_\mathrm{in},\mathbf{r'}_\mathrm{in}) \rangle  =  \langle H_{\textrm{in}}(\mathbf{r},\rin) H_\textrm{in}^*(\mathbf{r},\rpin) \rangle
%\end{equation}
To go further, a rough approximation is to assume an object of constant reflectivity in intensity:  $\langle |\gamma(\mathbf{r})|^2 \rangle=\gamma_0^2$.  The correlation length $r_F$ of the reflected wave-field then corresponds to the coherence length of the input focal spots. In the strong aberration regime, $r_F$ thus scales as the input diffraction limit $\delta^\textrm{0}_\textrm{in}$. The number $M_R$ of independent speckle grains in the focal plane then correspond to the number of input resolution cells mapping the object: 
\begin{equation}
\label{MR}
    M_R \sim ( {\Omega}/ \delta^\textrm{0}_\textrm{in})^2
\end{equation}
These theoretical derivations account for the spatial incoherence exhibited by the reflected wave-field in Fig.~\ref{fig1}C2.
This figure plots the auto-correlation function $\mathcal{C}_R(\Delta \mathbf{r}) $ of the reflected wave-field in the focal plane. $\mathcal{C}_R(\Delta \mathbf{r}) $ is computed by averaging the correlation matrix coefficients $C_R(\rin,\rpin)$ over couples $(\rin,\rpin)$ of same relative position $\Delta \mathbf{r}=\rin -\rpin$.

\subsection{Distortion matrix}

As highlighted by Fig.~\ref{fig1}C and demonstrated above, the reflection matrix displays a random feature both at its output and input in the strong aberration regime. Now we will show how the de-scan of the input focal spots in the focal plane reveals the spatial correlations between wave distortions.

In the general case (\textit{i.e} beyond the isoplanatic limit), the $\mathbf{D}$-matrix coefficients can be expressed as follows
\begin{equation}
\label{Dkr2}
D(\uout,\rin)= \int d\mathbf{r} \hat{H}_{\textrm{out}}(\uout,\mathbf{r})T_0(\uout,\mathbf{r}-\rin) \gamma(\mathbf{r}) H_{\textrm{in}}(\mathbf{r},\rin) \end{equation}
To investigate the spatial correlations of the distorted wave-field in the pupil plane, the correlation matrix $\mathbf{C_D}= {N^{-1}_\textrm{out}}\mathbf{D}^{\dag}\mathbf{D}$ should be considered. 
As the other correlation matrices, $\mathbf{C_D}$ can  be decomposed as the sum of a covariance matrix $\left \langle \mathbf{C_D}\right \rangle $ and a perturbation term $\delta  \mathbf{C_D} $ whose intensity is inversely proportional to the number, $N_D=(\delta_\textrm{in}/\delta_\textrm{out}^\textrm{0})^2$, of independent pupil speckle grains in the distorted wave-field (Eq.~\ref{ND}). 

For $N_D>>1$, $\mathbf{C_D}$ is shown to converge towards the covariance matrix $\langle \mathbf{C_D}\rangle$. Its coefficients can then be expressed as follows:
\begin{eqnarray}
\label{Cp1}
 C_D (\rin,\rpin)  & =& {N^{-1}_\textrm{out}} \int d\mathbf{r_1}  \int  d\mathbf{r_2}  
 H_\mathrm{in}(\mathbf{r_1},\rin) H_\mathrm{in}^*(\mathbf{r_2},\rpin) \gamma(\mathbf{r_1})  \gamma^*(\mathbf{r_2}) 
    \\ & & \times  \sum_{\uout} \langle \hat{H}_\textrm{out}(\uout,\mathbf{r_1})\hat{H}^*_\textrm{out}(\uout,\mathbf{r_2}) \rangle
   T_0(\uout,\mathbf{r_1}-\rin)T_0^*(\uout,\mathbf{r_2}-\rpin)\nonumber
   \end{eqnarray}
%This expression can be simplified by considering its ensemble average. A strong aberrating layer is indeed a source of randomness that may induce a self-averaging process.
%\begin{eqnarray}
%\label{Cp1}
 %C_D (\rin,\rpin)   =  \int d\mathbf{r_1}  \int  d\mathbf{r_2}  & & \left \langle \sum_{\mathbf{u}}  \hat{H}_\textrm{out}(\mathbf{u},\mathbf{r_1})\hat{H}^*_\textrm{out}(\mathbf{u},\mathbf{r_2}) \right \rangle \\
  %& & \times G_0(\uout,\mathbf{r_1}-\rin)G_0^*(\uout,\mathbf{r_2}-\rpin)\nonumber \\
 %& & \times \gamma(\mathbf{r_1})  \gamma^*(\mathbf{r_2})  H_\mathrm{in}(\mathbf{r_1},\rin) H_\mathrm{in}^*(\mathbf{r_2},\rpin) \nonumber
%\end{eqnarray}
Using Eqs.~\ref{G0}, \ref{F} and \ref{att}, the sum over $\uout$ in Eq.~\ref{Cp1} can be simplified as follows:\alex{
\begin{gather}
 {N^{-1}_\textrm{out}} \sum_{\uout}  \langle  \hat{H}_\textrm{out}(\uout,\mathbf{r_1})\hat{H}^*_\textrm{out}(\uout,\mathbf{r_2}) \rangle  T_0(\uout,\mathbf{r_1}-\rin)T_0^*(\uout,\mathbf{r_2}-\rpin) \nonumber \\
  = F (\mathbf{r_1},\mathbf{r_2}) \sum_{\uout}  \exp \left ( i \frac{2\pi}{\lambda f} \uout . (\mathbf{r_1}-\rin - \mathbf{r_2}+\rpin)\right ) \nonumber \\
 = F (\mathbf{r_1},\mathbf{r_2}) \delta (\mathbf{r_1}-\rin - \mathbf{r_2}+\rpin) 
 \label{random3}
\end{gather}}
If the statistical properties of the scattering medium are invariant by translation, then $F (\mathbf{r_1},\mathbf{r_2})=F (||\mathbf{r_1}-\mathbf{r_2}||)$. The spatial extension of the function $F$ directly yields the isoplanatic length $\ell_c$.
 The injection of Eq.~\ref{random3} into Eq.~\ref{Cp1} yields
\begin{equation}
\label{Cp2a}
 C_D (\rin,\rpin)  = F(\Delta r) \int d\mathbf{r}   \gamma(\mathbf{r}) \gamma^*(\mathbf{r}-\Delta \mathbf{r})  H_\mathrm{in}(\mathbf{r},\rin)H^*_\mathrm{in}(\mathbf{r}-\Delta \mathbf{r},\rpin) .
\end{equation}
with $\Delta \mathbf{r}=\rin-\rpin$ and $\Delta r=|\rin-\rpin|$. The factor $ F(\Delta r)$ requires that the correlation coefficients $ C_D (\rin,\rpin)$ cancel for points belonging to different IPs. The input PSFs can thus be  considered as locally invariant by translation, such that
$H_\mathrm{in}(\mathbf{r}-\rin+\rpin,\rpin) \simeq H_\mathrm{in}(\mathbf{r}-\rin)$. Equation~\ref{Cp2a} simplifies into
\begin{equation}
\label{Cp2b}
 C_D (\rin,\rpin)  \propto F(\Delta r)  \int d\mathbf{r} \gamma(\mathbf{r}) \gamma^*(\mathbf{r}-\Delta \mathbf{r}) | H_\mathrm{in}(\mathbf{r},\rin)|^2 ,
\end{equation}
To go further, we can assume that the width of the input focusing beam $\delta_\textrm{in}$ is larger than the characteristic fluctuation length $\ell_\gamma$ of the sample reflectivity:
\begin{equation}
\label{Cp2c}
 C_D  (\rin,\rpin)    \sim  F(\Delta r)  (\gamma \ast \gamma)(\Delta r).
\end{equation}
where the symbol $ \ast$ stands for the correlation product. Depending on the experimental conditions, the coherence length $d_F$ of the distorted wave-field can correspond to the correlation length $\ell_\gamma$ of the object's reflectivity or the isoplanatic length $\ell_c$ associated with the aberrating layer
\begin{equation}
\label{dF}
d_F=\mathrm{min} \left \lbrace \ell_c,\ell_\gamma \right \rbrace 
\end{equation} 
$d_F$ is thus always larger than the coherence length $r_F\sim \delta^\textrm{0}_\textrm{in}$ of the incoherent reflected wave-field (Eq.~\ref{C3}). The number $M_D$ of independent focal speckle grains for the distorted wave-field is given by  
\begin{equation}
    \label{MD}
    M_D=(\Omega/\ell_c)^2
\end{equation}
If $\ell_{\gamma}>\ell_c$, this number $M_D$ coincides with the number $(\Omega/\ell_c)^2$ of IPs contained by the object.  

These theoretical predictions confirm the experimental observations highlighted by Fig.~\ref{fig1}. Spatial correlations are drastically enhanced between the input entries of $\mathbf{D}$ (Fig.~\ref{fig1}D2) compared to $\mathbf{R}$ (Fig.~\ref{fig1}C2). Figure~\ref{fig1}D2 plots the auto-correlation function $\mathcal{C}_D(\Delta \mathbf{r}) $ of the distorted wave-field in the focal plane. This quantity is calculated by averaging the correlation matrix coefficients $C_D(\rin,\rpin)$ over couples $(\rin,\rpin)$ sharing the same relative position $\Delta \mathbf{r}=\rin -\rpin$.

Now, we show how the long-range correlations exhibited by $\mathbf{D}$ can be leveraged for overcoming the aberrations and retrieving an image of the object with a resolution close to the diffraction limit. 
%In the experiment depicted in Fig.~\ref{fig1}, the field-of-view is contained in a single isoplanatic patch ($F(\Delta r)=1$ across the field-of-view). The spatial correlation of the $\mathbf{D}$-matrix displayed in Fig.~\ref{fig1}D2 is thus dominated by the spatial frequency content of the object. In the experiments depicted in Figs.~\ref{fig3} and \ref{fig4}, the spatial correlations in $\mathbf{D}$ are, on the contrary, limited by the isoplanatic covariance function $F(\Delta r)$. In all cases, the distorted wave-fields exhibits much stronger spatial correlations than the reflected wave-field (Eq.~\ref{C2}). As we will see, those correlations in the focal plane can be predominant compared to those in the pupil plane (Eq.~\ref{C3}), especially in a strong aberration regime.

\section{\label{svd_supp}Singular value decomposition of the distortion matrix}

To take advantage of the correlations exhibited by the matrix $\mathbf{D}$, its SVD (Eq.~\ref{svd}) is shown to be an essential tool. It enables a decomposition of the FOI into IMs and an estimation of the transmission matrix $\mathbf{T}$ between the CCD surface and the focal plane. To provide a theoretical proof of this claim, the previous study of the correlation matrices $\mathbf{B_D}$ and $\mathbf{C_D}$ will be helpful. Their eigenvalue decomposition actually dictates the SVD of $\mathbf{D}$. Correlations in the focal plane are shown to predominate in the experiments depicted in the accompanying paper, but also, more generally, in optical microscopy. Strikingly, an exchange of role is noticed between the medium's reflectivity and the input PSF in the $\mathbf{D}$-matrix compared to the original $\mathbf{R}$-matrix. While the first singular vector of $\mathbf{R}$ yields the input PSF for a point-like reflector~\cite{Prada2003,popoff2}, the first singular vector of $\mathbf{D}$ directly yields the sample reflectivity for a point-like input focusing beam in an isoplanatic configuration. Beyond this analogy made between $\mathbf{R}$ and $\mathbf{D}$ in this asymptotic limit, a theoretical proof is then provided in the general case. We show how: (\textit{i}) the SVD of $\mathbf{D}$ allows a decomposition of the FOI into a set of IMs $\mathbf{V_p}$; (\textit{ii}) a coherent combination of the output eigenvectors $\mathbf{U_p}$ can lead to an estimator of the transmission matrix $\mathbf{T}$.   

\subsection{Eigenvalue decomposition of the correlation matrices}

The SVD of $\mathbf{D}$ (Eq.~\ref{svd}) can be directly deduced from the eigenvalue decompositions of its correlation matrices $\mathbf{B_D}$ and $\mathbf{C_D}$. The latter ones can actually be written as follows
\begin{equation}
\label{BD}
\mathbf{B_D}=\mathbf{U}\mathbf{\Sigma}^2\mathbf{U^{\dag}}
\end{equation}
and 
\begin{equation}
\label{CR}
\mathbf{C_D}=\mathbf{V}\mathbf{\Sigma}^2\mathbf{V^{\dag}}.
\end{equation}
or, in terms of matrix coefficients,
\begin{equation}
\label{svd3}
B_D(\uout,\upout)=\sum_{p=1}^{ {N_\textrm{in}}} \sigma_p^2 U_p(\uout) U^*_p(\upout).
\end{equation}
and
\begin{equation}
\label{svd4}
C_D(\rin,\rpin)=\sum_{p=1}^{{N_\textrm{in}}} \sigma_p^2 V_p(\rin) V^*_p(\rpin).
\end{equation}
The eigenvalues of $\mathbf{B_D}$ and $\mathbf{C_D}$ are the square of the singular values $\sigma_p$; their eigenvectors, $\mathbf{U_p}$ and $\mathbf{V_p}$, are the output and input singular vectors, respectively. The SVD of $\mathbf{D}$ is dictated either by the correlations between its lines or columns. To know which ones dominate over the other, the analytical expressions of the correlation matrices, $\mathbf{B_D}  $ and $ \mathbf{C_D}$, should be investigated (see Eqs.~\ref{C3} and \ref{Cp2c}). 

If the reflectivity of the object was fully random, \textit{i.e} $\langle \gamma(\mathbf{r}) \ast \gamma(\mathbf{r}) \rangle =\delta (\mathbf{r})$, the correlation matrix $ \mathbf{C_D}$ (Eq.~\ref{Cp2c}) would be diagonal. This means that the columns of $\mathbf{D}$ would be fully uncorrelated. On the contrary, output correlations would subsist in $\mathbf{D}$ as they only depend on the spatial extension of the input focal spot (Eq.~\ref{thetaD}). In this random speckle regime, the SVD of $\mathbf{D}$ is dominated by its correlations in the pupil plane and the analysis of $\mathbf{D}$ should rather be restricted to a FOI containing a single IP. This regime has been recently investigated in medical ultrasound imaging~\cite{Lambert2020} where scattering is often due to a random distribution of unresolved scatterers. 

In optical microscopy, biological tissues induce a strong forward scattering: The involved scatterers display a characteristic length $\ell_{\gamma}$ larger than the wavelength. The auto-correlation of the sample reflectivity can span over several IPs especially at large depths. In this forward scattering regime, correlations of the distorted wave-field in the focal plane may dominate over its far-field correlations.  

To know if this is the case, one can compare the number of independent speckle grains, $N_D$ and $M_D$, in the pupil and focal planes, respectively. The correlation degree in each plane is actually inversely  proportional to this number. Correlations in the focal plane will thus dominate if $N_D>M_D$. The latter condition is fulfilled in a strong aberration regime for which the number of output resolution cells mapping each aberrated focal spot, $N_D =(\delta_\mathrm{in} /\delta_\textrm{out}^\textrm{0})^2$, is larger than the number of IPs mapping the object surface, $M_D=(\Omega/\ell_c)^2$. This condition is checked in the experiments of the accompanying paper. For instance, in the experiment depicted in Fig.~\ref{fig4}, $N_D\sim 500 $ while $M_D \sim 10$.

\subsection{Analogy with iterative time reversal}

Now that the conditions for a domination of correlations in the focal plane have been derived, we now study the singular vectors of $\mathbf{D}$. To that aim, an analogy with iterative time reversal is first explored to give a physical intuition of the SVD of $\mathbf{D}$.

If we compare the analytical expressions of the correlation matrices $\mathbf{C_R}$ (Eq.~\ref{C2}) and $\mathbf{C_D}$ (Eq.~\ref{Cp2a}), we can notice an exchange of the role between the medium reflectivity $\gamma$ and the input PSF $H_\textrm{in}$. While $\mathbf{C_R}$ corresponds to a static object scanned by a moving illuminating beam (Fig.\ref{fig3}A), $\mathbf{C_D}$ corresponds to a static focused beam illuminating a moving object (Fig.\ref{fig3}B). In the isoplanatic limit, the distortion matrix $\mathbf{D}$ (Eq.~\ref{Cp2a}) is thus equivalent to a virtual reflection matrix associated with: (\textit{i}) a coherent reflector of scattering distribution $|H_\textrm{in}(\mathbf{r})|^2$ (located on the optical axis and at the focal plane); (\textit{ii}) a virtual focusing beam associated with the PSF $\gamma(\rin+\mathbf{r})$. As shown by iterative time reversal experiments~\cite{prada1994eigenmodes,Prada2003}, the reflection matrix is of rank 1 for a point-like scatterer, and its first input singular vector $\mathbf{V_1}$ shall directly yield the virtual input PSF~\cite{badon2016smart}. By analogy, for a point-like input focusing beam, the $\mathbf{D}$-matrix shall be also of rank 1 and its first input singular vector $\mathbf{V_1}$ shall directly provide the medium reflectivity $\gamma(\rin)$. Interestingly, the SVD of $\mathbf{D}$ should therefore unscramble aberrations and sample reflectivity. However, this qualitative analysis has been made under strong hypotheses: the isoplanatic limit and a point-like input focusing beam. In the following, we make the problem more complex by first going beyond the isoplanatic limit and then by considering the finite size of the input focusing beams. 
%For analytical tractability, we will now investigate the fictitious case of a nearly point-like input focusing beam, before further tackling the reality, \texit{i.e} strongly aberrated input focusing beams.  

\subsection{Isoplanatic modes}

Let us first assume a point-like input focusing beam, $H_\textrm{in}(\mathbf{r},\rin)=|H_\textrm{in}(\rin,\rin)|^2\delta(\mathbf{r}-\rin)$, beyond the isoplanatic limit. Equation \ref{Cp2a} becomes
\begin{equation}
\label{Cpoint}
 C_D (\rin,\rpin)  \propto F(\Delta r) \gamma(\rin) \gamma^*(\rpin) H_\mathrm{in}(\rin,\rin) H^*_\mathrm{in}(\rpin,\rpin).
\end{equation}
A full-field intensity image $\mathcal{F}(\rin)$ of the sample reflectivity can be retrieved by considering the diagonal of $\mathbf{C_D}$:
 \begin{equation}
\label{image}
\mathcal{F}(\rin)=C_D(\rin,\rin)= |\gamma(\rin)|^2 |H_\textrm{in}(\rin,\rin)|^2
\end{equation}
$\mathcal{F}(\rin)$ can be a satisfying estimator of the sample reflectivity, $|\gamma(\rin)|^2$. However, the input focusing beam intensity $H_\textrm{in}(\rin,\rin)$ pollutes the full-field image. The latter term can be detrimental to imaging since it gives rise to a fluctuating contrast across the focal plane. Moreover, experimental noise and diffusive multiple scattering can still degrade the image. At last, we may want to have access to the amplitude and phase of the reflectivity rather than only its square norm. For all these reasons, the singular value decomposition of $\mathbf{D}$ (Eq.\ref{svd}), or equivalently, the eigenvalue decomposition of $\mathbf{C_D}$(Eq.\ref{svd4}) is decisive. In the general case, the correlation function $F(\Delta r)$ (Eq.~\ref{FF}) governs the eigenvalue decomposition of $\mathbf{C_D}$. The ratio between the object surface ${\Omega^2}$ and the isoplanatic area $\ell_c^2$ yields the effective rank $M_D=(\Omega/\ell_c)^2$ of $ \mathbf{C_D}$.
This rank scales as the number of IPs that fit in the object. The input eigenvectors $\mathbf{V_p}$ can be derived by solving a second order Fredholm equation with Hermitian kernel~\cite{Ghanem}. An analytical solution can be found for certain analytical form of the correlation function $F(\Delta r)$ (Eq.~\ref{FF}). For instance, a sinc kernel imply 3D prolate sphero\"{i}dal eigenfunctions~\cite{robert2009}; a Gaussian covariance function leads to Hermite-Gaussian eigenmodes~\cite{aubry2006}; exponential or triangular kernels yields cosine and sine eigenfunctions~\cite{Ghanem}. A general trend is that the spatial frequency content of the eigenvectors increases with their rank.

The identification of Eqs.~\ref{svd4} and \ref{Cpoint} leads to the following equality:
\begin{equation}
\label{Cp2}
 \sum_{p=1}^{M_D} \sigma_p V_p(\rin) = H_\mathrm{in}(\rin,\rin)\gamma(\rin)
\end{equation}
A coherent combination of the $M_D$ first eigenvectors $\mathbf{V_p}$ can yield the amplitude and phase of the reflectivity but the result is still polluted by the input illumination beam $H_\textrm{in}(\rin,\rin)$. 
In practice, aberrations at the input can be corrected through the same process by exchanging input and output, \textit{i.e} by projecting the data in the pupil plane at the input and in the focal plane at the ouput. In the experiments depicted in the accompanying paper, the sparse illumination scheme makes the input basis incomplete and the spatial sampling insufficient. The image should thus be built from the output to benefit from the excellent resolution with which the field is recorded by the CCD camera. To do so, Eq.~\ref{Cp2} can be used to prove that the coherent combination of output singular vectors $\mathbf{U_c}=\sum_{p=1}^{M_D} \mathbf{U_p}$ (Eq.~\ref{svd}) perfectly compensate for the output aberration matrix $\mathbf{\hat{H}_\textrm{out}}$. To that aim, let us apply the transpose conjugate $\mathbf{U_c^{\dag}}$ to the output of the matrix $\mathbf{D}$ (Eq.\ref{Dkr2}). It comes:
\begin{equation}
  \int d\mathbf{r}  \sum_{\mathbf{u}} U_c^*(\mathbf{u}) H_{\textrm{out}}(\mathbf{u},\mathbf{r})T_0(\mathbf{u},\mathbf{r}-\rin) \gamma(\mathbf{r}) H_{\textrm{in}}(\mathbf{r},\rin) \nonumber
= H_\mathrm{in}(\rin,\rin)\gamma(\rin)
\end{equation}
This last equality is valid only and only if 
\begin{equation}
   \sum_{\mathbf{u}} U_c^*(\mathbf{u}) H_{\textrm{out}}(\mathbf{u},\mathbf{r})T_0(\mathbf{u},\mathbf{r}-\rin)=\delta (\mathbf{r}-\rin)
\end{equation}
which, under the matrix formalism, can be rewritten as 
\begin{equation}
(\mathbf{U_c}\circ \mathbf{T_0})^{\dag} \mathbf{T}= \mathbb{I}
\end{equation}
The matrix $\mathbf{\hat{T}}=(\mathbf{U_c}\circ \mathbf{T_0})$ is an estimator of the transmission matrix $\mathbf{T}$. The application of its transpose conjugate, $\mathbf{\hat{T}}^{\dag}$ enables a perfect compensation for the aberrations contained in the transmission matrix $\mathbf{T}$. To obtain a diffraction-limited image of the object, the matrix  $\mathbf{\hat{T}}^{\dag}$ should be directly applied to the output of the matrix $\mathbf{R}$ (Eq.\ref{PSF} of the accompanying paper). This operation leads to a focused matrix $\mathbf{R_F}$ whose coefficients can be expressed as
\begin{equation}
\label{RFF}
    R_F(\rout,\rin)=\gamma(\rout) H_\textrm{in}(\rout,\rin)
\end{equation}
This matrix consists in an Hadamard product between the reflectivity of the focal plane at its output and the input focusing matrix. In other words, aberrations are corrected at the output but subsists at the input. Hence the resulting confocal image built from the diagonal of $\mathbf{R_F}$ suffers from the same issue:
\begin{equation}
\label{IF}
  \mathcal{I} (\rout)= R_F(\rout,\rout) = \gamma(\rout) H_\textrm{in}(\rout,\rout)  
\end{equation}
It is a relying estimator of the object's reflectivity $\gamma(\rout)$, but modulated by the amplitude and phase of the input illumination $H_\textrm{in}(\rout,\rout)$. To reduce this detrimental effect on the image contrast, one can consider a full-field image integrated over all input focusing beams (see Eq.~\ref{F} of the accompanying paper) or an adaptive confocal image integrating over a numerical pinhole (see Eq.\ref{I} of the accompanying paper). This integration over $\rin$ allows us to smooth the modulation of the image induced by the input focusing beams.

\subsection{Finite size of the input PSF}

All these theoretical developments have been made by considering a point-like input focused beam. This is, of course, not true in reality. The input focusing beam gives rise to a virtual coherent reflector of finite size $\delta_\textrm{in}$. The issue we want to address is the impact of this size on the SVD of $\mathbf{D}$. Assuming incoherent input focusing beams (Eq.~\ref{random}), Eq.~\ref{Cp2b} can be rewritten as follows in the isoplanatic limit (Eqs.~\ref{iso1}-\ref{iso2}):
\begin{equation}
 C_D (\rin,\rpin)  \propto  \left ( \int d\mathbf{r} \gamma(\mathbf{r}) H_\mathrm{in}(\mathbf{r}-\rin) \right ) \times  \left ( \int d\mathbf{r'} \gamma(\mathbf{r'}) H_\mathrm{in}(\mathbf{r'}-\rpin) \right )^*,
\end{equation}
By confronting this last equation with the eigenvalue decomposition of Eq.~\ref{svd4}, it turns out that the distortion matrix $\mathbf{D}$ is of rank 1 and that its input singular vector $\mathbf{V_1}$ can be expressed as
\begin{equation}
\label{V1a}
V_1(\rin)=\int d\mathbf{r} \gamma(\mathbf{r}) H_\mathrm{in}(\mathbf{r}-\rin)= [\gamma \circledast H_\mathrm{in}](\rin).
\end{equation}
where the symbol $\circledast$ stands for the convolution product. Albeit independent from output aberrations, $\mathbf{V_1}$ is nevertheless a convolution product between the object's reflectivity and the input PSF $H_\mathrm{in}$ (see Fig~\ref{fig3}C of the accompanying paper). The output singular vector $\mathbf{U_1}$ can be deduced from $\mathbf{V_1}$ through the following matrix product:
\begin{equation}
\sigma_1 \mathbf{U_1}=\mathbf{D}\mathbf{V_1}.
\end{equation} 
Injecting Eq.~\ref{Dkr} and Eq.~\ref{V1a} into this last equation yields the following expression for the coefficients of $\mathbf{U_1}$
\begin{equation*}
   \sigma_1 \mathbf{U_1} (\uout)= \hat{H}_\textrm{out}(\uout)  \int d \mathbf{r} \int   d\mathbf{r'} \sum_{\rin} T_0(\uout,\mathbf{r}-\rin) \gamma(\mathbf{r}) \gamma(\mathbf{r'})
 H_\textrm{in}(\mathbf{r}-\rin) H_\textrm{in}(\mathbf{r'}-\rin) 
\end{equation*}
For a large number of resolution cells in the FOI, $\mathbf{U_1}$ will converge towards its ensemble average, such that
\begin{equation*}
   \sigma_1 \mathbf{U_1} (\uout)= \hat{H}_\textrm{out}(\uout)  \int d \mathbf{r} \int   d\mathbf{r'} \sum_{\rin} T_0(\uout,\mathbf{r}-\rin) \gamma(\mathbf{r}) \gamma(\mathbf{r'})
 \left \langle H_\textrm{in}(\mathbf{r}-\rin) H_\textrm{in}(\mathbf{r'}-\rin) \right \rangle 
\end{equation*}
In a strong aberration regime (Eq.~\ref{random}), the last equation can be rewritten as follows
\begin{equation*}
   \sigma_1 \mathbf{U_1} (\uout)= \hat{H}_\textrm{out}(\uout)  \int d \mathbf{r} |\gamma(\mathbf{r})|^2 \sum_{\rin} T_0(\uout,\mathbf{r}-\rin) 
 \left \langle \left | H_\textrm{in}(\mathbf{r}-\rin) \right |^2 \right \rangle 
\end{equation*}
Using the expression of the free-space transmission coefficients $T_0(\uout,\rin)$ (Eq.~\ref{G0}), it finally turns out that
\begin{equation}
 U_1(\uout) \propto \hat{H}_\textrm{out}(\uout) \left [ \hat{H}_\textrm{in} \ast \hat{H}_\textrm{in} \right ](\uout).
\end{equation}
and 
\begin{equation}
 \sigma_1 \propto  \int d \mathbf{r} |\gamma(\mathbf{r})|^2 .
\end{equation}
While the singular value $\sigma_1$ yields the object's reflectivity \alex{integrated} over the associated isoplanatic patch (here the FOI), the vector $\mathbf{U_1}$ corresponds to the aberration output transmittance $\hat{H}_\textrm{out}$ modulated by the autocorrelation function of the aberration input transmittance $\hat{H}_\textrm{in}$ (see Fig~\ref{fig3}C of the accompanying paper). This last term tends to limit the angular aperture of the singular vector $\mathbf{U_1}$ by the coherence angle of the input aberration $\hat{H}_\textrm{in}$. To circumvent that issue, the trick is to consider only the phase of the first singular vector $\mathbf{U_1}$. Indeed, if we make the realistic hypothesis of a real and positive autocorrelation function $\hat{H}_\textrm{in} \ast \hat{H}_\textrm{in}$, the normalized vector $\mathbf{\tilde{U}_1}$ is then given by
\begin{equation}
\label{U1a}
{\tilde{U}_1}(\uout)=\exp \left ( j \mbox{arg} \left \lbrace U_1(\uout) \right \rbrace \right )=\hat{H}(\uout)
\end{equation}
A novel input vector $\mathbf{\tilde{V}_1}$ can then be retrieved through the matrix product: 
\begin{equation}
\mathbf{\tilde{U}_1}^{\dag}\mathbf{D}= \mathbf{\hat{V}_1}.
\end{equation}
Injecting the expression of $\mathbf{\tilde{U}_1}$ (Eq.~\ref{U1a}) and $\mathbf{D}$ (Eq.~\ref{Dkr}), the following expression can be retrieved for $\mathbf{\hat{V}_1}$ in the isoplanatic limit:
\begin{equation}
  \hat{V}_1 (\rin) = H_\mathrm{in}(\rin,\rin)\gamma(\rin)
\end{equation}
If we compare this last equation with Eq.~\ref{V1a}, the normalization of $\mathbf{U_1}$ allows us to virtually reduce the size of the input focusing beam (see Fig~\ref{fig3}D of the accompanying paper). The  matrix $\mathbf{\hat{T}}=(\mathbf{\tilde{U}_1}\circ \mathbf{T_0})$ is then a satisfying estimator of the transmission matrix $\mathbf{T}$ in the isoplanatic limit. The application of its transpose conjugate, $\mathbf{\hat{T}}^{\dag}$, allows a perfect compensation for the aberrations contained in the transmission matrix $\mathbf{T}$. A diffraction-limited image of the object can be obtained by applying the matrix $\mathbf{\hat{T}}^{\dag}$ to the output of matrix $\mathbf{R}$ (Eq.~\ref{PSF}).  

\subsection{General case}

In the general case (\textit{i.e} beyond the isoplanatic limit), the same method can be employed to virtually reduce the size of the focal spot over each IM. The corresponding singular vectors should be normalized: $\mathbf{\tilde{U}_p}=\exp \left ( j \mbox{arg} \left \lbrace \mathbf{U_p} \right \rbrace \right )$. The application of their transpose conjugate to the $\mathbf{R}$-matrix should lead to an optimal aberration correction over each IM at the output. One open question is whether these output singular vectors can be combined coherently or not, such that $\mathbf{\tilde{U}_c}=\sum_p \mathbf{\tilde{U}_p}$. In the present work, this coherent combination does not provide better results than an incoherent summation of each IM image $\mathcal{I}_p$ (Eq.\ref{I2}). This is because an incoherent sum of $\mathbf{R_F}$-matrix coefficients at the input is required to smooth the modulation of the image by $H_\textrm{in}$ (Eqs.\ref{F}-\ref{I}).

\newpage 

\section*{Supplementary Figures}

\begin{figure}[ht!]
\center
\includegraphics[width=0.9\textwidth]{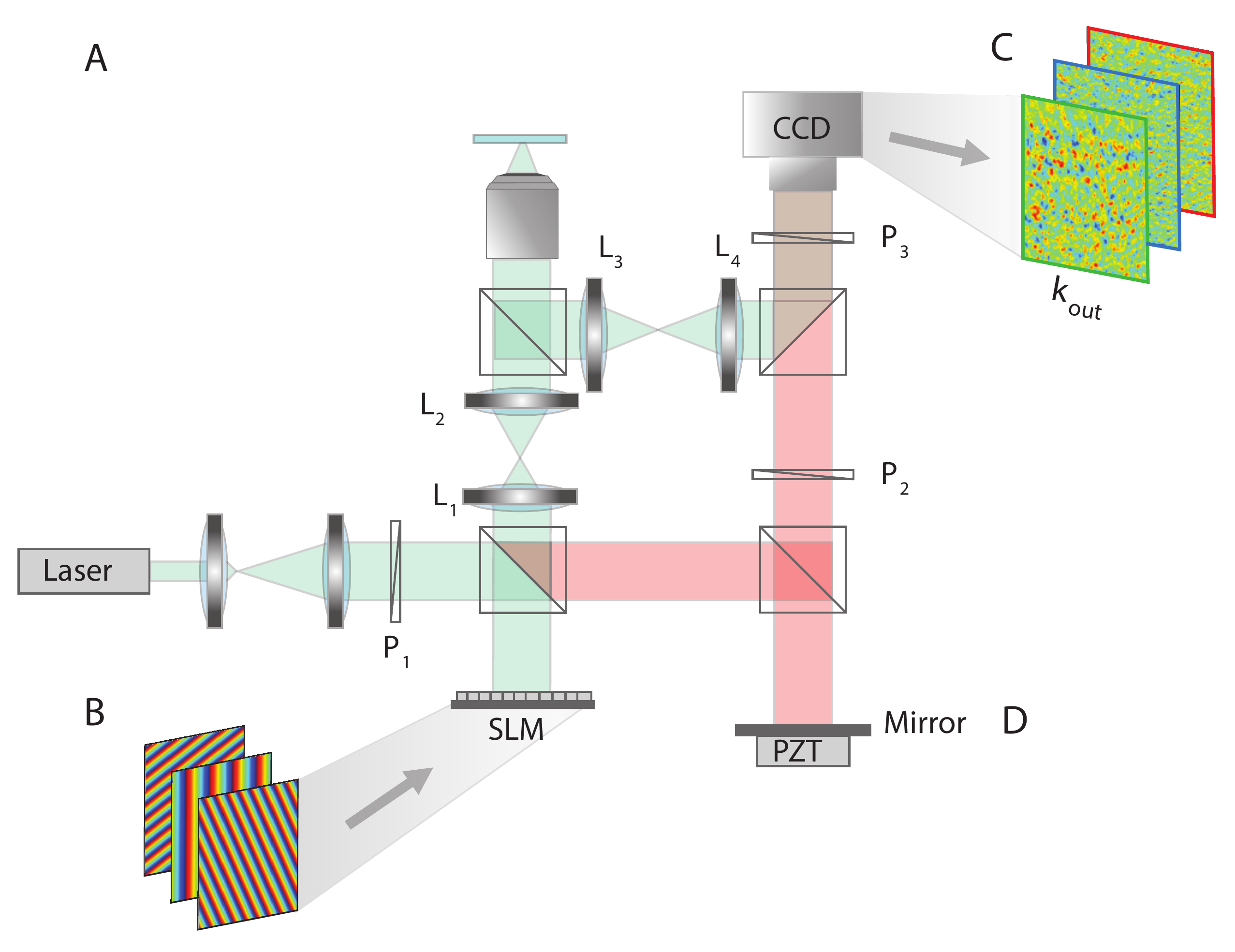}
\caption{\label{S1}\textbf{Measuring the time-gated reflection matrix}. Experimental set up: P: polarizer, MO: microscope objective, BS: beam splitter, PBS: polarized beam splitter, SLM : spatial light modulator, PZT: piezo phase shifter, M: Mirror. A femtosecond laser beam (center wavelength: 810 nm, bandwidth: 40 nm) is shaped by an SLM acting as a diffraction grating. A set of incident plane waves is thus emitted from the SLM and focused at a different position in the focal plane of an immersion MO (NA=0.8). The backscattered wave-field is collected through the same MO and interferes with a reference beam on a CCD camera. The latter one is conjugated with the back focal plane of the MO. The amplitude and phase of the wave-field is recorded by phase shifting interferometry \cite{popoff2010measuring}. The time of flight $t$ is controlled by the length of the interferometric arm and is matched with the position of the focal plane. For each input focusing point $\mathbf{r_\textrm{in}}$, a reflected wave-field is recorded in the pupil plane and stored along a column vector in the matrix $\mathbf{R})
=[R(\mathbf{u_\textrm{out}},\mathbf{r_\textrm{in}})]$ .}
\end{figure}

\newpage

\begin{figure}[ht!]
\center
\includegraphics[width=0.7\textwidth]{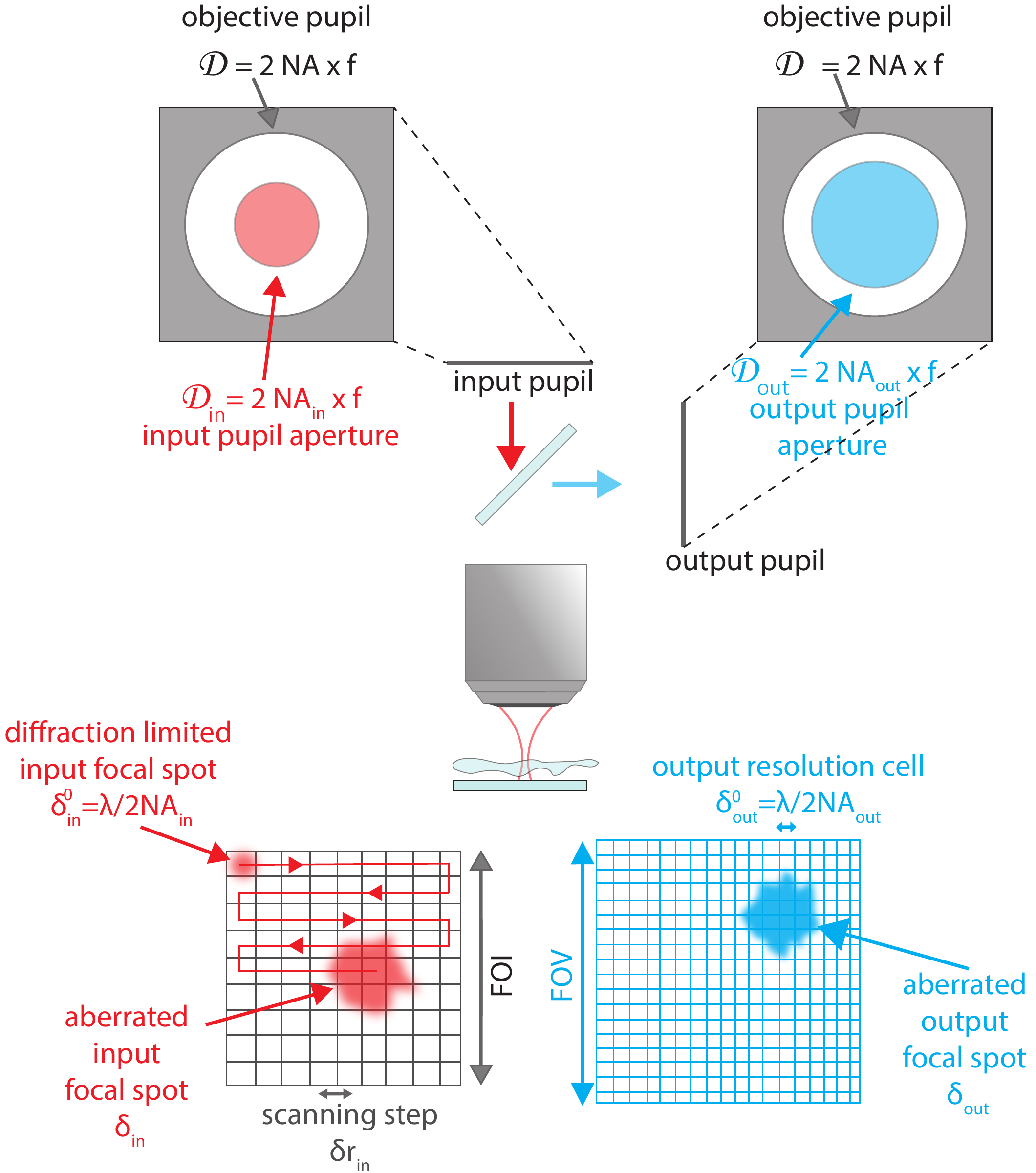}
\caption{\label{S7} \textbf{Conjugation relationships between the pupil, focal and imaging planes}. The distortion matrix connects the focal plane of the MO with the output pupil plane. This figure illustrates the various parameters involved in the different planes of the system. Both input and output pupil planes are ultimately limited by the MO edges. The input pupil $\mathcal{D}_\textrm{in}$ is even more limited because the illumination beam underfills the objective pupil $\mathcal{D}$. It results in a reduced NA denoted $\text{NA}_\textrm{in}$. In turn, the size of the input focal spot in the image plane is given by $\delta_\textrm{in}^\textrm{0}=\lambda/2\text{NA}_\textrm{in}$ if there is no aberration and $\delta_\textrm{in}$ in the general case. 
In this focal plane, the field-of-illumination depends on the scanning step (spatial sampling), denoted as $\delta r_\textrm{in}$, and the number of measurements $N_\textrm{in}$. In reflection, the output pupil $\mathcal{D}_\textrm{out}$ is also smaller than the total objective pupil $\mathcal{D}$ due to the limited surface of the detector but larger than the input pupil $\mathcal{D}_\textrm{in}$. The resolution of the image is thus governed by the output resolution cell $\delta_\textrm{out}^\textrm{0}=\lambda/2\text{NA}_{in}$.}
\end{figure}

%\newpage
%\begin{figure}[h!]
%\center
%\includegraphics[width=\textwidth]{figureS8.pdf}
%\caption{\label{S8}\textbf{Aberration correction for a defocus}. (A) Schematic of the experiment. A resolution target (USAF 1951) is positioned below the focus of the MO. (B) Original full-field image. (C) Matrix image constructed with the first eigenstate of D. (D) Plot of the normalized singular values of the distortion matrix. (E) Phase of the first singular vector $\mathbf{U_1}$.}
%\end{figure}

\newpage

\begin{figure}[ht!]
\center
\includegraphics[width=1\textwidth]{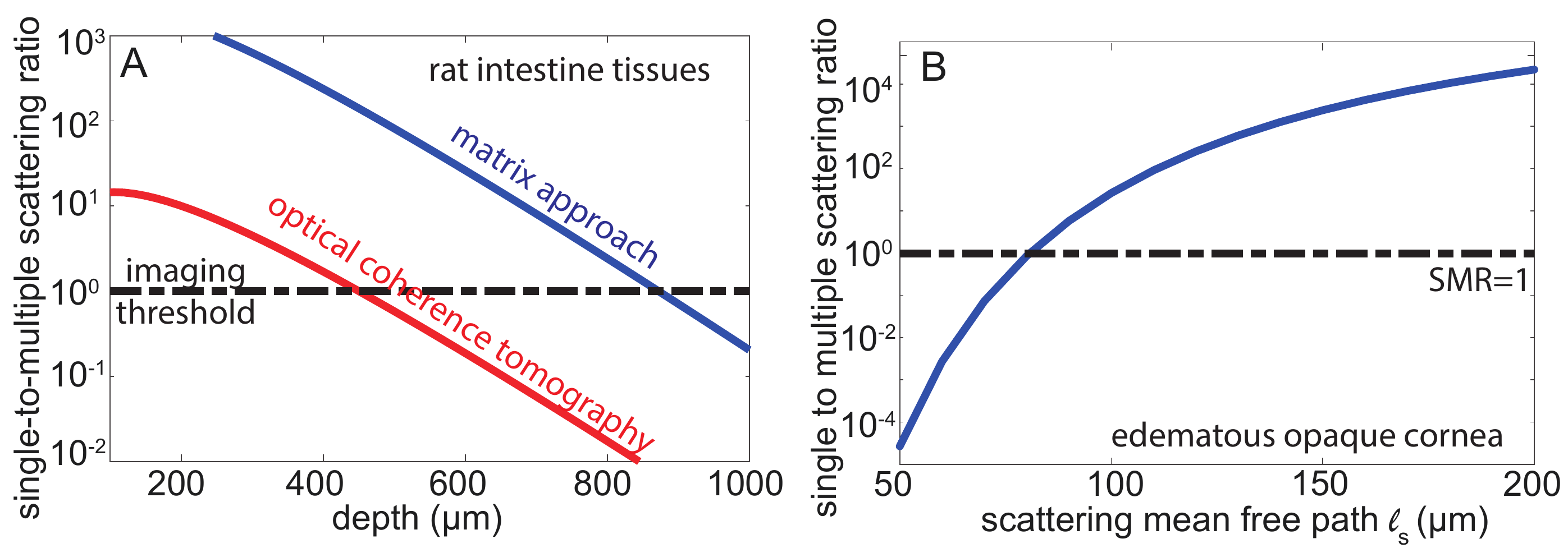}
\caption{\label{suppMS}\textbf{Predicting the single-to-multiple scattering ratio in biological tissues.} (\textbf{A}) SMR as a function of depth for the imaging experiment through the rat intestinal tissue (see Figs.~\ref{fig1} and \ref{fig2}). The red curve (before aberration correction) is plotted for a Strehl ratio $\mathcal{S}=3\times 10^{-3}$, while the blue curve (after matrix correction) corresponds to $\mathcal{S}=1.1\times 10^{-2}$. The detection threshold yields an imaging depth limit of $\sim$ 450 $\mu$m for conventional OCT and 900 $\mu$m for our matrix approach. (\textbf{B}) SMR as a function of the scattering mean free path $\ell_s$ for the cornea imaging experiment (see Fig.~\ref{fig5}). A $\textrm{SMR}$ of 1 is obtained for a scattering mean free path $\ell_s\sim$ 80 $\mu$m. In both panels, the theoretical curves are built by considering the model described in Ref.~\cite{badon2017multiple} and the experimental parameters described in the paper. Note also that the y-axis is in log-scale.}

\end{figure}

\newpage

\begin{figure}[ht!]
\center
\includegraphics[width=10cm]{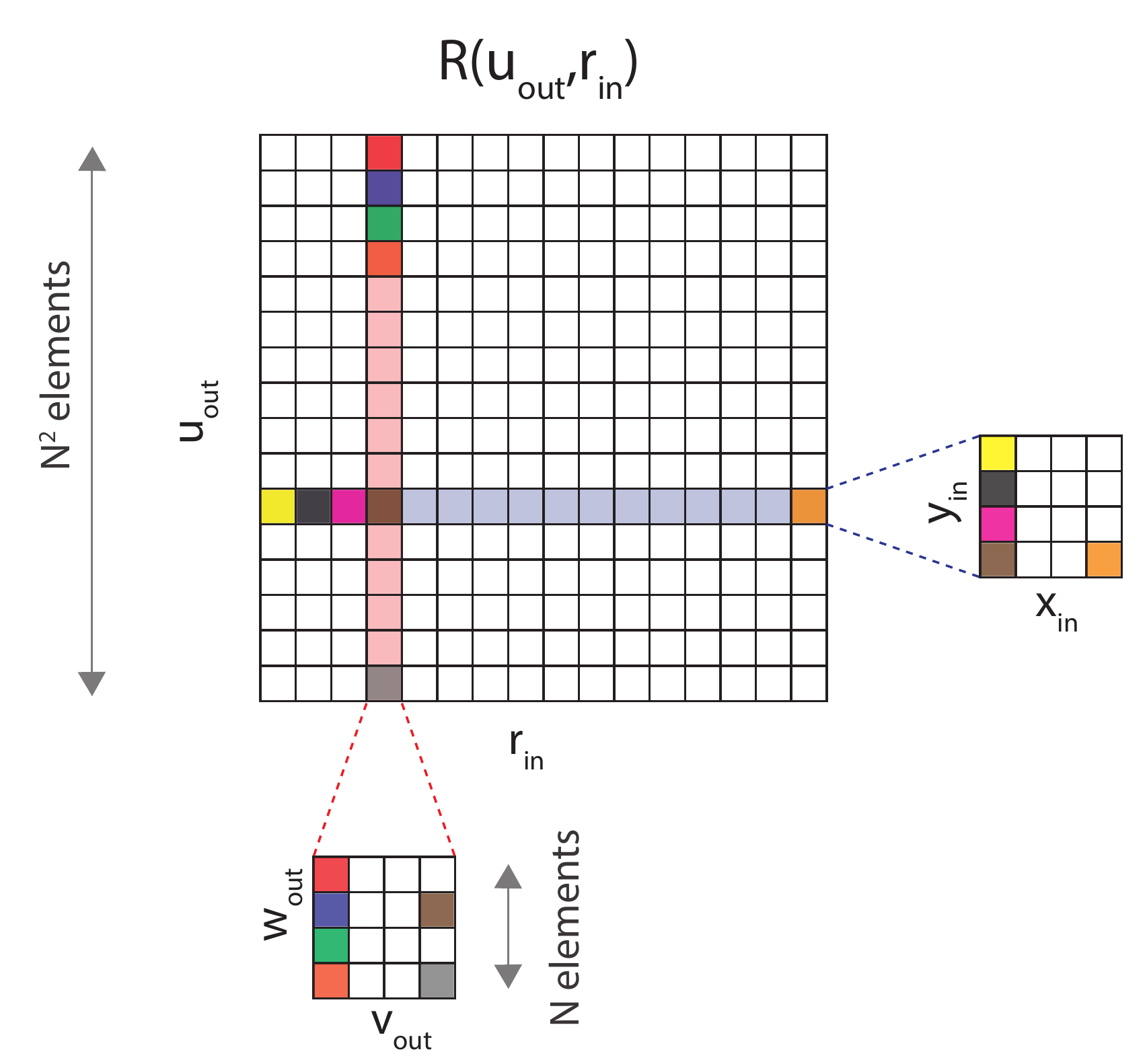}
\caption{\label{S2}\textbf{Building the reflection matrix $\mathbf{R}$}. For each focused illumination at $\mathbf{r_\textrm{in}}=(x_\textrm{in},y_\textrm{in})$, the reflected wave-field $\psi_{\mathbf{r_\textrm{in}}}(v_\textrm{out},w_\textrm{out})$ is recorded in the pupil plane by each pixel of the CCD camera whose position is denoted by the vector $\mathbf{u_\textrm{out}}=(v_\textrm{out},w_\textrm{out})$ . Each wave-field is concatenated and stored along a column vector. This set of column vectors forms the reflection matrix $\mathbf{R} = [R(\mathbf{u_\textrm{out}},\mathbf{r_\textrm{in}})]$, such that $R(\mathbf{u_\textrm{out}},\mathbf{r_\textrm{in}})=\psi_{\mathbf{r_\textrm{in}}}(v_\textrm{out},w_\textrm{out})$. }
\end{figure}

\newpage

\begin{figure}[h!]
\center
\includegraphics[width=\textwidth]{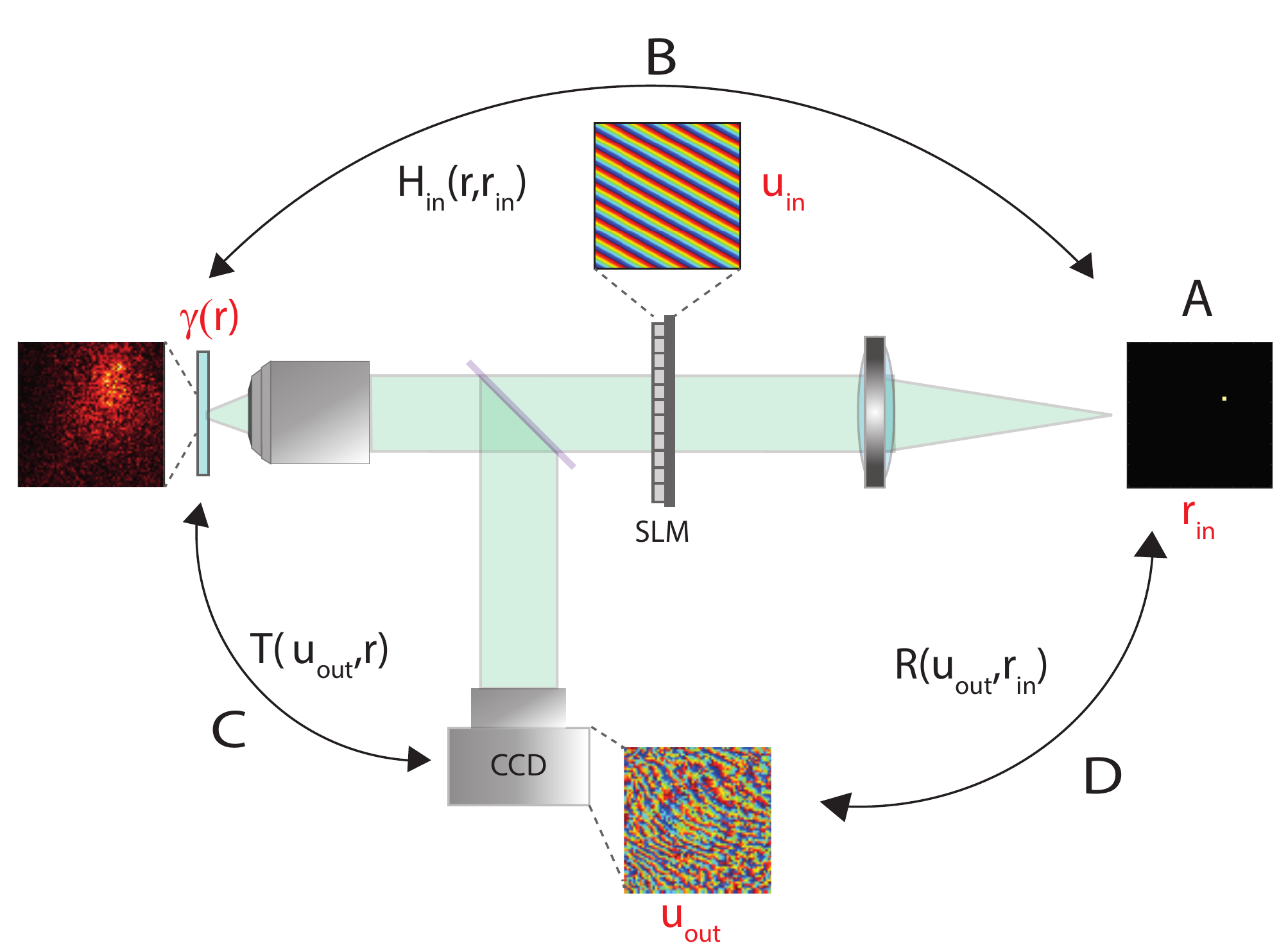}
\caption{\label{S3}\textbf{Modeling light propagation from the virtual source plane to the output pupil plane.} The reflection matrix $\mathbf{R}$ contains the impulse responses $R(\mathbf{u_\textrm{out}},\mathbf{r_\textrm{in}})$ between each virtual source point $\mathbf{r_\textrm{in}}$ and each CCD pixel $\mathbf{u_\textrm{out}}$ in the output pupil plane. ($\textbf{A}$) The virtual source point $\mathbf{r_\textrm{in}}$ is produced by each transverse mode shaped by the SLM in the input pupil plane. ($\textbf{B}$) The propagation between the virtual source plane and the focal plane of the MO can be modelled by the input focusing matrix $\mathbf{H_\textrm{in}}=[H_\textrm{in}(\mathbf{r},\mathbf{r_\textrm{in}})]$ whose columns correspond to each input focal spot in the sample plane for each incident focusing point $\mathbf{r_\textrm{in}}$. ($\textbf{C}$) The return trip of the wave from the sample to the CCD camera is modeled by the transmission matrix $\mathbf{T}=[T(\mathbf{u_\textrm{out}},\mathbf{r})]$ that connects each point $\mathbf{r}$ in the focal plane to each pixel $\mathbf{u_\textrm{out}}$ of the CCD camera. ($\textbf{D}$) Finally, based on these propagation matrices and the sample reflectivity matrix $\mathbf{\Gamma}$, the reflection matrix $\mathbf{R}$ can be simply expressed as the matrix product of these three matrices (Eq.\ref{int_abe}).}
\end{figure}

\newpage

\section*{Supplementary tables}

{\small
\begin{table*}[h!]
\center
\begin{tabular}{c|l}
variable   &  definition\\
\hline
\hline
$\lambda$ & wavelength\\
$n$ & optical index  \\
$g$ & anisotropy factor \\
$\ell_s$ & scattering mean free path \\
$L$ & thickness of the scattering layer \\
$d$ & distance between the aberrating layer and the focal plane \\
$f$ & focal length of the microscope objective \\
$\Omega$ & size of the field-of-illumination \\
$\rin$ / $\rout$ & position vector in the input focusing / output pupil planes \\
$\mathbf{r}$ & position vector in the focal plane of the microscope objective \\
$\uout$ & position vector in the output pupil plane \\
$N_\textrm{in}$ / $N_\textrm{out}$  & number of input focusing beams / pixels in the output pupil plane \\ 
$\mathcal{D}_\textrm{in}$ / $\mathcal{D}_\textrm{out}$ & input / output pupil aperture \\
$\mathrm{NA}_\textrm{in}$ / $\mathrm{NA}_\textrm{out}$  & input / output numerical aperture \\
$\delta r_\textrm{in}$ / $\delta u_\textrm{out}$ & spatial sampling in the input focusing / output pupil planes \\
$\delta_\textrm{in}$ / $\delta_\textrm{out}$ & characteristic width of the input/output point spread functions \\
$\delta^\textrm{0}_\textrm{in}$ / $\delta^\textrm{0}_\textrm{out}$ & diffraction limit resolution at input/ouput \\
$r_P$ / $r_F$ & correlation length of the reflected wave-field in the output pupil / input focusing plane \\
$d_P$ / $d_F$  & correlation length of the distorted wave-field in the output pupil / input focusing plane\\
$\ell_\gamma$ & charateristic correlation length of the sample's reflectivity \\
$\ell_c$ & characteristic size of an isoplanatic patch \\
$M_D$ / $N_D$ & number of independent speckle grains for $\mathbf{D}$ in the input focusing / output pupil planes \\
$M_R$ / $N_R$ & number of independent speckle grains for $\mathbf{R}$ in the input focusing / output pupil planes \\
$\mathcal{S}_0$ / $\mathcal{S}_p$ / $\mathcal{S'}_p$  & Strehl ratios: initial /final / weighted values   \\
$\sigma_p$ / $\tilde{\sigma}_p$  & singular values of $\mathbf{D}$: raw / normalized \\
$\mathcal{H}(\tilde{\sigma}_p)$ & Shannon entropy of singular values \\
SMR & single-to-multiple scattering ratio \\
\end{tabular}
 \caption{\label{table}\textbf{Glossary of the variables used in this study.}}
\end{table*}}

\newpage

{\small
\begin{table*}[h!]
\center
\begin{tabular}{l|l}
matrix   &  definition\\
\hline
\hline
$\mathbf{R}=[R(\uout,\rin)]$ & dual reflection matrix \\
$\mathbf{R_0}=[R_0(\rout,\rin)]$ & focused reflection matrix built from $\mathbf{T_0}$ \\
$\mathbf{R_p}=[R_p(\rout,\rin)]$ & focused reflection matrix built from $\mathbf{\hat{T}_p}$ \\
$\mathbf{R_F}=[R_F(\rout,\rin)]$ & focused reflection matrix built from $\mathbf{\hat{T}}$ \\
$\mathbf{T}=[T(\uout,\mathbf{r})]$ & transmission matrix  \\
$\mathbf{T_0}=[T_0(\uout,\mathbf{r})]$ & free-space transmission matrix \\
$\mathbf{\hat{T}}=[\hat{T}(\uout,\rout)]$ & estimator of the transmission matrix  \\
$\mathbf{\hat{T}_p}=[\hat{T}_p(\uout,\mathbf{r})]$ & estimator of the transmission matrix built from $\mathbf{U_p}$  \\
$\mathbf{\Gamma}=[\gamma(\mathbf{r})]$ & sample's reflectivity matrix \\
$\mathbf{\Gamma_D}=[\gamma_D(\mathbf{r})]$ & virtual scatterer reflectivity matrix  \\
$\mathbf{H}_\textrm{in}=[{H}_\textrm{in}(\mathbf{r},\rin)]$ &  input focusing matrix \\
$\mathbf{D}=[D(\uout,\rin)]$ & distortion matrix \\
$\mathbf{U_p}=[U_p(\uout)]$ & output singular vector of $\mathbf{D}$ \\
$\mathbf{\tilde{U}_p}=[\tilde{U}_p(\uout)]$ & normalized output singular vector of $\mathbf{D}$ \\
$\mathbf{V_p}=[{V}_p(\rin)]$ & input singular vector of $\mathbf{D}$ \\
$\mathbf{B_R}=[B_R(\uout,\upout)]$ & correlation matrix of $\mathbf{R}$ in the pupil plane \\
$\langle \mathbf{B_R} \rangle =[\mathcal{B}_R(\Delta \mathbf{u})]$ & covariance matrix of $\mathbf{R}$ in the puil plane \\
$\mathbf{B_D}=[B_D(\uout,\upout)]$ & correlation matrix of $\mathbf{D}$ in the pupil plane \\
$\langle \mathbf{B_D} \rangle =[\mathcal{B}_D(\Delta \mathbf{u})]$ & covariance matrix of $\mathbf{D}$ in the pupil plane \\
$\mathbf{C_R}=[C_R(\uout,\upout)]$ & correlation matrix of $\mathbf{R}$ in the focal plane \\
$\langle \mathbf{C_R} \rangle =[\mathcal{C}_R(\Delta \mathbf{r})]$ & covariance matrix of $\mathbf{R}$ in the focal plane \\
$\mathbf{C_D}=[C_D(\uout,\upout)]$ & correlation matrix of $\mathbf{D}$ in the focal plane \\
$\langle \mathbf{C_D} \rangle =[\mathcal{C}_D(\Delta \mathbf{r})]$ & covariance matrix of $\mathbf{D}$ in the focal plane \\
$ \mathbf{\hat{H}}_\textrm{out}  =[{\hat{H}}_\textrm{out} (\uout,\mathbf{r})]$ & aberration matrix \\
$\mathbf{F}=[ F (\mathbf{r},\mathbf{r'})]$ & correlation matrix of $\mathbf{\hat{H}}_\textrm{out} $ in the focal plane \\
\end{tabular}
 \caption{\label{table}\textbf{Glossary of the matrices used in this study.}}
\end{table*}}

\end{document}